# Energy and Scaling Limits of Phase-Change Memory

Rivka-Galya Nir-Harwood[1], Asir I. Khan[2,3], Emanuel Ber[1], Efrat Ordan[1], Keren Stern[1], Kye L. Okabe[2], Nicolás Wainstein[1], Eilam Yalon[1*], and Eric Pop[2,4,5,6**]

[1]*Viterbi faculty of Electrical & Computer Engineering, Technion - Israel Institute of Technology, Haifa, 32000, Israel.*
[2]*Department of Electrical Engineering, Stanford University, Stanford, CA 94305, USA.*
[3]*Lawrence Berkeley National Laboratory, Berkeley, CA 94720, USA.*
[4]*Department of Materials Science & Engineering, Stanford University, Stanford, CA 94305, USA.*
[5]*Department of Applied Physics, Stanford University, Stanford, CA 94305, USA.*
[6]*Precourt Institute for Energy, Stanford University, Stanford, CA 94305, USA.*

[*]*E-mail: eilamy@technion.ac.il*
[**]*E-mail: epop@stanford.edu*

**Phase change memory (PCM) relies on a reversible transition between amorphous and crystalline states of a material, and stands as a promising candidate for next-generation, energy-efficient data storage and neuromorphic hardware. Here, we review key innovations that have driven PCM technology to achieve energy consumption down to only tens of femtojoules per bit, and could further advance it closer to its fundamental limits. Because PCM switching is induced thermally, we highlight improvements in energy-efficiency through two primary strategies: by minimizing the active phase change material region to sub-10 nm dimensions, and by enhancing heat confinement within PCM devices to reduce thermal dissipation into the surrounding environment. While the theoretical limits could reach single attojoules per cubic nanometer of memory material, realizing these limits in practice is significantly constrained by electrical and thermal parasitics, particularly at contacts and interfaces.**

## Contents

## 1. Introduction

In an era dominated by data-abundant and power-hungry computing [1] – spanning social networks, e-commerce, artificial intelligence platforms and large language models – energy efficiency has emerged as a critical global concern [2], [3], [4], [5], [6]. To address this challenge, the development of densely integrated ultra-low power and low-energy electronics is essential. Among these, non-volatile data storage technologies such as phase-change memory (PCM) [7], resistive RAM (RRAM) [8], conductive bridge RAM (CBRAM) [9], and spin-transfer torque magnetic RAM (STT-MRAM) [10] are at the forefront. These technologies hold the promise of significantly reducing energy consumption, particularly within analog in-memory computing and neuromorphic hardware, by minimizing the energy wasted during data transfer between memory and computing units [11]. Furthermore, in computing systems, non-volatile memory plays a crucial role in reducing idle energy consumption and simplifying memory access.

PCM offers unique opportunities for low-power data storage as discussed in this review, in addition to fast access speed (<60 ns), extended endurance (>$10^9$), long retention time (>10 years) and high density (<$4F^2$) [7], [12], [13]. PCM is based on phase-change materials, most commonly ternary and binary chalcogenide compounds like $Ge_2Sb_2Te_5$ (GST), $Sb_2Te_3$, and GeTe, a class of materials originally conceived in the late 1960s by Stanford R. Ovshinsky [14]. Phase change materials can be reversibly switched between two main phases (states): an ordered, most often (poly)crystalline phase, and a disordered, amorphous phase, with low and high electrical resistivity, respectively. The optical properties between these phases also exhibit notable differences, with PCMs finding commercial applications in rewritable optical discs since the late 1980s [15]. Nonetheless, data stored optically faces size limitations due to the diffraction limit, typically in the range of microns to a few hundred nanometers. This review concentrates on the electrical aspects of PCM technology.

In the past two decades PCM has been proposed as a potential replacement for non-volatile Flash [16] and for bridging the performance gap between (volatile) memory and (long-term) data storage [17]. The data (logic '1' or '0') are encoded by the resistivity of the chalcogenide material, which for most phase change materials can change by over four orders of magnitude between the amorphous and crystalline phase [18] (**Figure 1a**). With such large on/off ratio, PCM devices have also demonstrated multibit operation by exploiting intermediate resistance states, further improving storage density [19], [20], and standing as a main candidate for neuromorphic computing hardware [21], [22], [23], [24]. In the latter case, the electrical conductance can represent synaptic weights in artificial neural networks (ANN), e.g. in crossbar arrays for matrix-vector multiplication [21], [22], [23], [24], [25], [26] and as the membrane potential of a neuron in spiking artificial neural networks (SNNs) [27], [28].

The memory state can be reversibly switched with fast (nanosecond scale) electrical pulses (**Figure 1b)**, which induce phase change through Joule heating. Crystallization (an operation called "set") is achieved by heating the PCM above its crystallization temperature ($T_C$) for sufficient time so that crystallization takes place (typically > 100 ns). Amorphization (an operation called "reset") is achieved by a melt-quench process, heating the PCM above its melting temperature ($T_M$ ~ 600 °C for GST) and cooling it faster than the crystallization rate to "freeze" the material in a disordered state. The rapid cooling is obtained by a short trailing edge (fall time) of the applied pulse, which is critical for the melt-quench process. Measurement of the fast-switching energy requires probing the transient current (with sub-100-ns timescale). Such fast-transient measurements are non-trivial, explaining the relatively scarce data reporting switching energy in PCM.

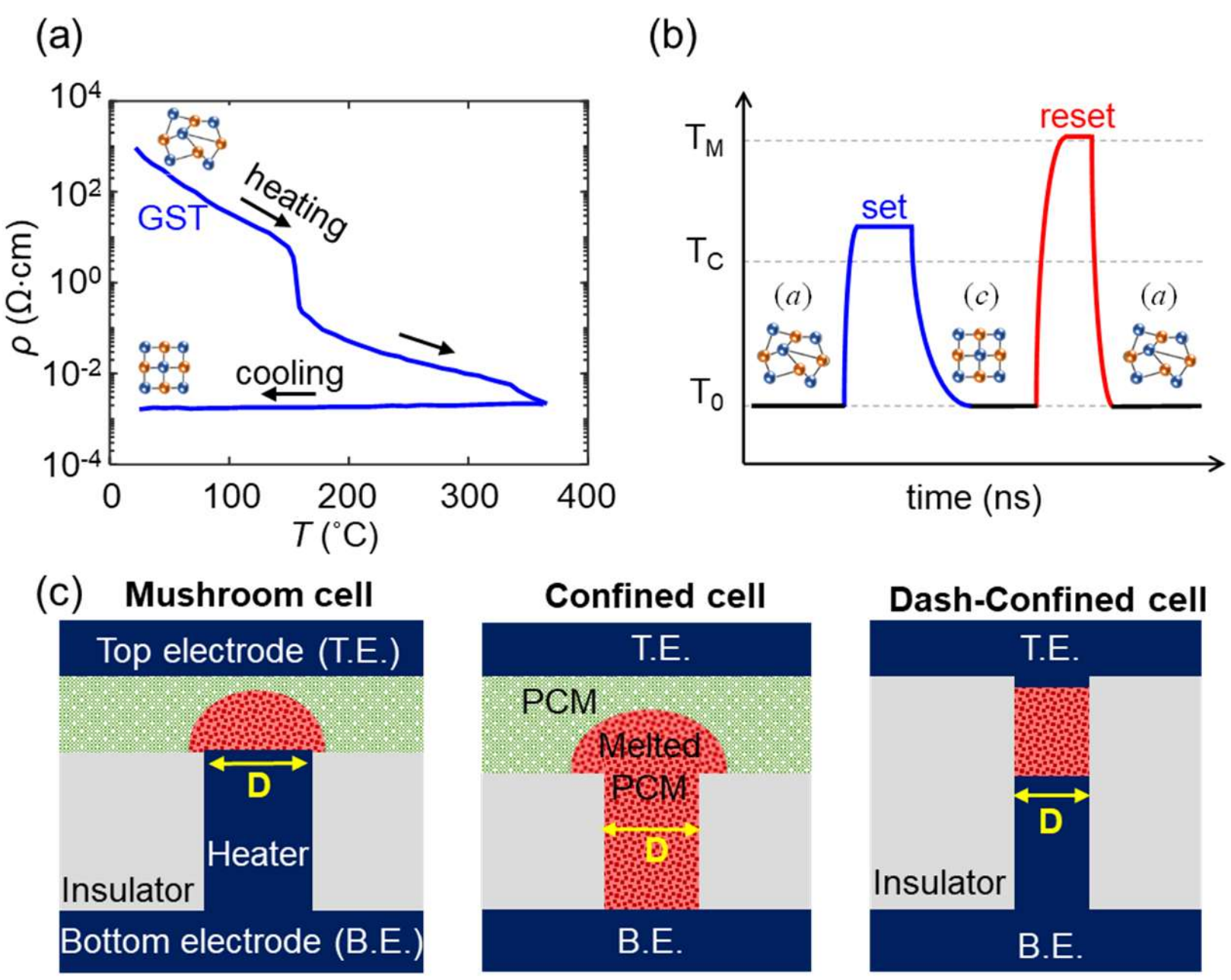

*Figure 1* **PCM operation.** (a) Resistivity of a common phase-change material, $Ge_2Sb_2Te_5$ (GST), as a function of temperature, starting with an as-deposited, 50-nm thick amorphous film *[18]*. Arrows indicate direction of heating and cooling. Adapted from S. Raoux *et al.*, J. Appl. Phys., 2007, with permission from AIP Publishing. (b) Programming cycles of PCM. Heating beyond a certain temperature ($T_C$ ~ 160 ºC for GST) crystallizes the phase change material. Heating to the melting temperature ($T_M$ ~ 600 ºC for GST) and rapidly quenching the material returns it to the amorphous state. (c) Schematic cross-sectional device structure of three typical PCM cells: mushroom, confined, and dash confined. The critical dimension of the bottom electrode or confinement is marked as "*D*" (diameter), and the active PCM region during device operation is shown in red.

The narrow current path, such as a small-area bottom electrode (BE) heater or confined PCM, determines its critical dimension (*D* in **Figure 1c** and Supporting Information section S1). The resistance change of the PCM cell from low (crystalline phase) to high (amorphous phase) occurs when the active volume, defined by the critical dimension, is amorphized, effectively "blocking" the current. Hence, the reset current is typically directly proportional to the active area (~$D^2$) and the switching energy to the active volume.

Therefore, it is evident that the simplest way to reduce the switching current and energy is by scaling down the critical dimension $D$ of the PCM cell.

Although PCM switching energy scales with size, it is still relatively high compared to alternative emerging non-volatile memory (NVM) technologies. This is primarily due to the large reset energy, which typically ranges from hundreds of fJ to hundreds of pJ, arising from the high currents required for Joule heating of the phase-change material above its melting temperature ($T_M$). The set operation also incurs a non-negligible energy cost—spanning from tens of fJ to hundreds of pJ—mainly because it requires relatively long programming pulses to allow for crystallization, typically exceeding 100 ns. An inherent trade-off between fast set operation and long retention time, governed by the crystallization temperature: lower crystallization temperatures enable faster set speeds but at the expense of reduced retention time [7]. Since the reset operation typically requires higher energy than the set operation, this review primarily focuses on the reset process.

**Figure 2a** compares the programming energy of emerging NVM technologies vs. their active contact area. These memory devices are two-terminal structures, illustrated schematically in **Figure 2b**, where a switching element (with cross-sectional area $A$ and diameter $D$) is sandwiched between top and bottom electrodes. The dashed lines represent the energy required to charge and discharge the interconnects: the top dashed line corresponds to an entire 1k × 1k array ($2^{20}$ bits), while the bottom dashed line represents the energy for a single random bit or a pair of metal interconnects. The interconnect energy calculations are detailed in Supporting Information Section S2. It is evident that alternative technologies can be programmed with energies comparable to or even lower than the interconnect charging energy of random 1 bit. In the case of RRAM and CBRAM the switching is localized in nanometer-sized filaments, enabling low energy operation (*e.g.* sub-pJ), often at the expense of a lower on/off ratio than PCM [29]. Because the filament formation is confined and is independent of the device contact dimension, such filamentary memories do not exhibit scaling of energy with cell dimensions.

Scaling PCM offers many advantages including higher data density, improved endurance [30], higher speed [30], [31], [32], and lower switching current [31], [32], [33], [34], [35]. This review focuses primarily on the switching energy aspect of scaling. For more information on other scaling improvements, we refer to other excellent reviews [7], [12], [36], [37].

This manuscript reviews two approaches to reducing the switching energy of PCM: 1) scaling the cell dimensions and, 2) improving heat confinement to reduce energy loss to the surroundings. The second approach, improving heat confinement, is further divided into two strategies: 2a) thermal engineering of barriers or surrounding materials to increase thermal resistance and, 2b) shortening the switching pulse width (PW) to reduce heat diffusion time.

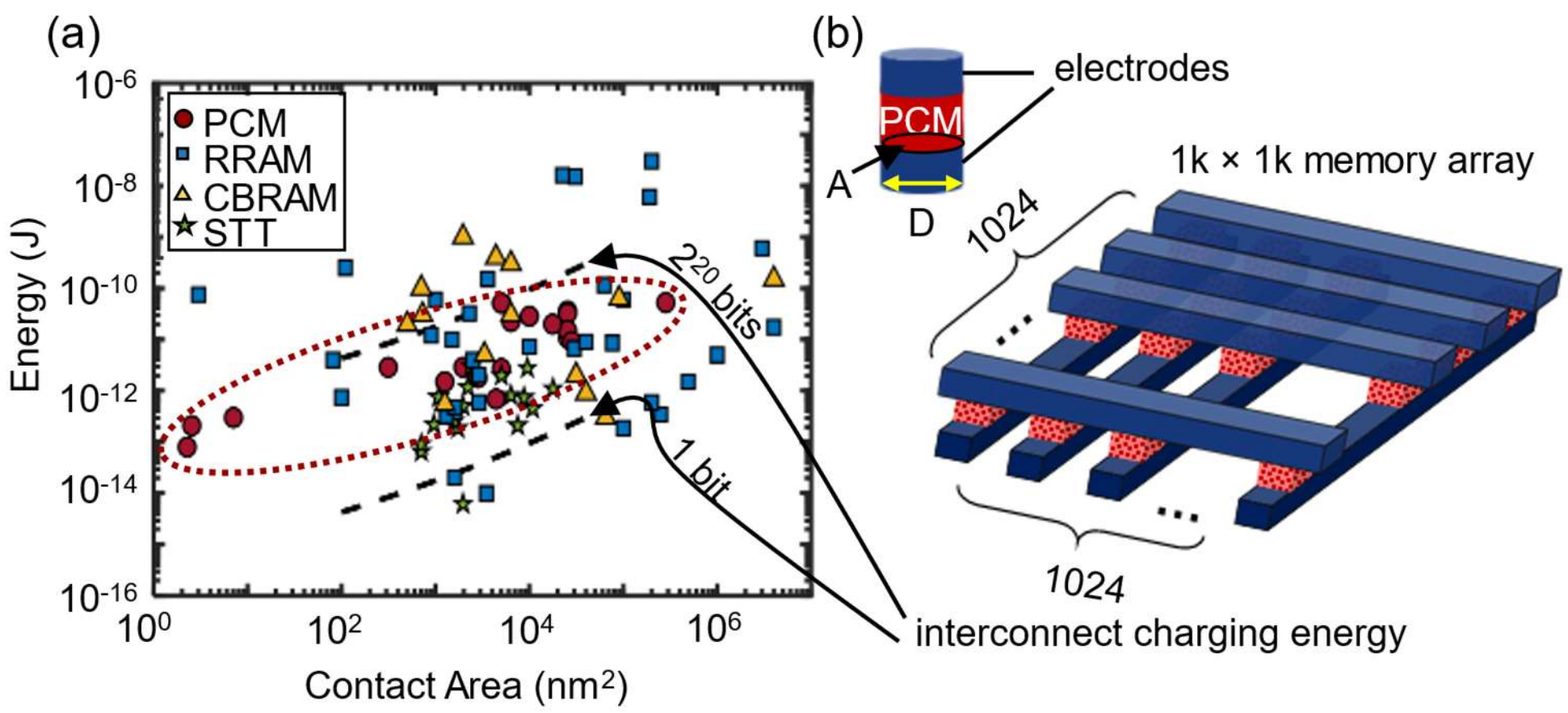

*Figure 2* **Benchmarking programming energy.** (a) Comparison of write energy vs. contact area for four emerging non-volatile memory technologies *[38]* : PCM (red circles), RRAM (blue squares), CBRAM (yellow triangles), and STT-MRAM (green stars). Note that for PCM only reset energy values are shown for they are typically higher than set energy values. (b) Schematic of a single cell and a 1k × 1k memory array. The energy required to charge the entire array (1k × 1k) and a single random bit are shown in (a) for comparison (black dashed lines). Reducing the device programming energy below that of a single bit line offers no practical benefit. PCM can only reset using short pulses (or short fall times) due to the melt-quench process. PCM programming energy scales with contact area (non-filamentary behavior), illustrated by the dashed red ellipse as a guide to the eye.

This review is organized as follows: first, in Section 2, we discuss scaling solutions that enable lower energy consumption. This includes an exploration of carbon nanotubes (CNTs), graphene, and filamentary electrodes for minimizing device contact areas. It is important to note that as device dimensions approach sub-100 nm lengths (lateral) or thicknesses (vertical), the contacts and interfaces increasingly dominate their electrical and thermal behavior. Therefore, Section 3 delves into this dominance and elaborates on superlattice (SL) devices.

In Section 4 we present the two strategies mentioned earlier for enhancing heat confinement in PCM: increasing thermal resistance and shortening the PW. Both strategies aim to reduce the energy dissipated in heating the PCM surrounding. For heat confinement through higher thermal resistance, we review improvements using thermal insulators, such as the out-of-plane direction in van der Waals (vdW) materials (*e.g.*, molybdenum disulfide, $MoS_2$) and amorphous materials (*e.g.*, tantalum oxide, $Ta_2O_5$) as interfacial layers. Additionally, we discuss how sub-nanosecond reset pulses significantly enhance heat confinement and achieve substantial energy efficiency improvements. These short reset pulses highlight the limitation of set energy, which becomes comparable to the reset energy due to the extended crystallization time. Consequently, we also address improvements in set speed in Section 5. Finally, in this last section, we explore the ultimate energy and scaling limits of PCM technology.

## 2. PCM Scaling

## a) PCM with Carbon Nanomaterials

In this section we discuss scaling and energy improvements in PCM devices enabled by carbon nanomaterials, specifically carbon nanotubes (CNTs) and graphene used as electrodes [39]. Conventional PCM cells are typically contacted with metal electrodes, such as TiN, which also serve as "heaters." An important demonstration occurred in 2011, showing that replacing metal contacts with CNT electrodes enabled PCM cells to be scaled down to single-nanometer dimensions (**Figure 3a**), achieving reset currents of ~5 μA and switching energies of a few hundred fJ per bit [40]. Subsequently, self-aligned lateral CNT devices showed reversible switching using current pulses (**Figure 3b**), with reset currents below 1.6 μA and energies below 40 fJ per bit, approximately two orders of magnitude lower than state-of-the-art PCM devices at the time [41].

These advancements were made possible by recognizing that CNTs can carry extremely large current densities (~GA/cm$^2$) at diameters of a few nm without perils of electromigration, and that they can form atomically sharp contacts to PCM. This represented a crucial insight, demonstrating that PCM is a data-storage technology that can continue to benefit from aggressive scaling (**Figure 3c**), unlike charge-based memories, where quantum leakage currents fundamentally limit scaling below ~10 nm [42]. Moreover, these results highlighted that the fundamental scaling limits of PCM had not yet been reached. Subsequent work further confirmed that CNT electrodes are highly effective in addressing nanometer-scale PCM cells, leading to substantial reductions in programming current and energy [43].

Building on these concepts, graphene has emerged as a closely related carbon-based electrode material that preserves many of the advantages of CNTs while offering different integration and scalability trade-offs. Like CNTs, graphene is made of the same sp$^2$-bonded carbon lattice and is atomically thin (~0.34 nm). Although graphene is difficult to pattern into ribbons as narrow as a single CNT, it is easier to process by conventional methods, and is potentially more compatible with CMOS technology and even flexible electronics [44]. This makes graphene an appealing candidate as an edge electrode for PCM devices, as demonstrated by Behnam *et al.* [45]. The power consumption of the graphene-edge-contact PCM devices was approximately an order of magnitude greater than those with CNT electrodes (consistent with the larger contact area between the graphene edges and the phase change material). Nevertheless, programming currents remained in the single μA range, with threshold voltages around ~3 V, and median off/on resistance ratios of ~100. While scalable, the variability and reliability of such devices still require improvement, particularly through better control of the PCM interfaces with both the substrate and graphene contacts [45].

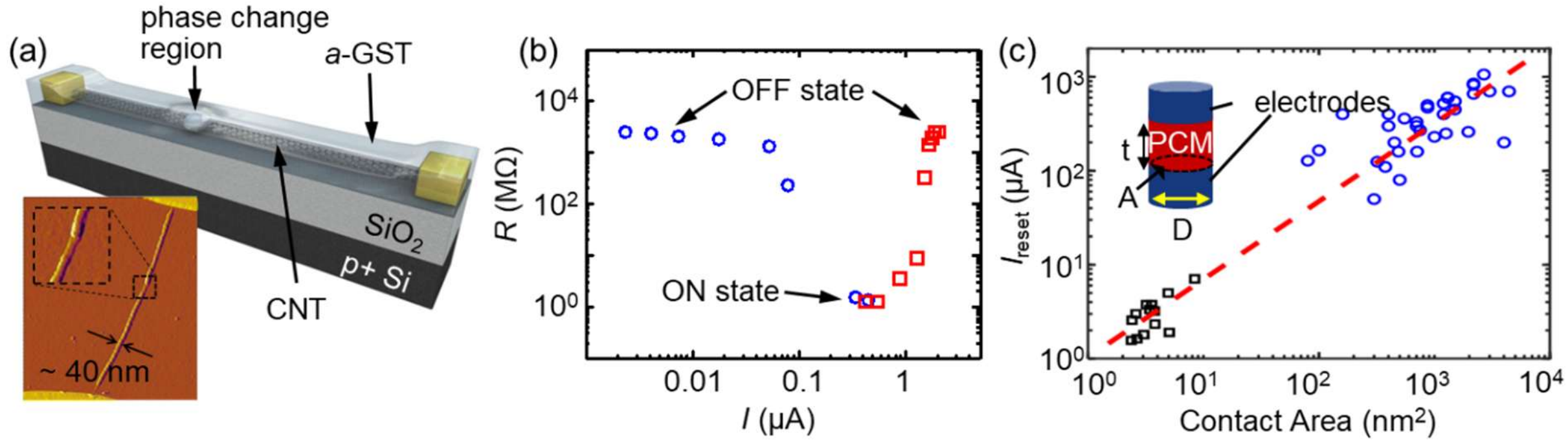

*Figure* 3 **Low-power PCM device with CNT electrodes.** (a) Device schematics *[40]* and an atomic force microscopy image of the device *[41]*. (b) Set (blue) and reset (red) switching with current pulses lower than 2 μA *[41]*. (c) PCM reset current scaling vs. electrode contact area. Blue dots are from the literature *[7], [33]* and black squares are PCM devices with CNT electrodes. Inset shows idealized device schematic. The figures are adapted from *[40], [41], [46]* with permission.

## b) Filamentary PCM

Recent studies have demonstrated reduced PCM contact area using conductive filaments with diameters of just a few nm, formed electrically [47] [48]. A similar technique was employed in the early 2000s to fabricate sub-lithography constrictions for PCM devices [49], [50]. The concept aligns with the approach discussed previously, aiming to realize nanoscale PCM cells, but with the key distinction that the nanometer-scale contact is formed electrically. For instance, a self-confined $SiTe_x$ nano-filament was electrically formed by injecting tellurium atoms into an amorphous silicon (a-Si) layer (**Figure 4a**) [48]. The effective device diameter was ~5.5 nm with a reset current of ~60 μA (**Figure 4b**). By reducing the amount of tellurium in the $SiTe_x$ filament, a lower reset current of around 10 μA was achieved. The devices exhibited switching times of 150 and 20 ns for set and reset respectively, an on/off ratio greater than 100, endurance exceeding $10^5$, and energy consumption below 10 pJ for both set and reset. Although this work achieved a relatively low reset current, the operating voltage exceeded 3 V. It is important to benchmark all relevant device properties to better understand the trade-offs involved.

An earlier study [47] used monolayer graphene and $MoS_2$, separately, to reduce switching energy by creating a vdW interface between BE and PCM, to achieve better thermal confinement. The vdW material, with its large cross-plane thermal resistance [51], [52], [53], helps block heat dissipation from the PCM to the BE. While reduced switching energy was achieved, this was primarily due to the formation of a filamentary contact through an oxidized TiN layer on the bottom electrode (**Figure 4c**). In this case, the reset current was independent of the BE diameter; instead, the effective bottom interface area was determined by the filament [47]. The combination of reduced effective contact area and thermal confinement provided by the $MoS_2$ monolayer resulted in the most significant reduction in switching current and power (**Figure 4d**).

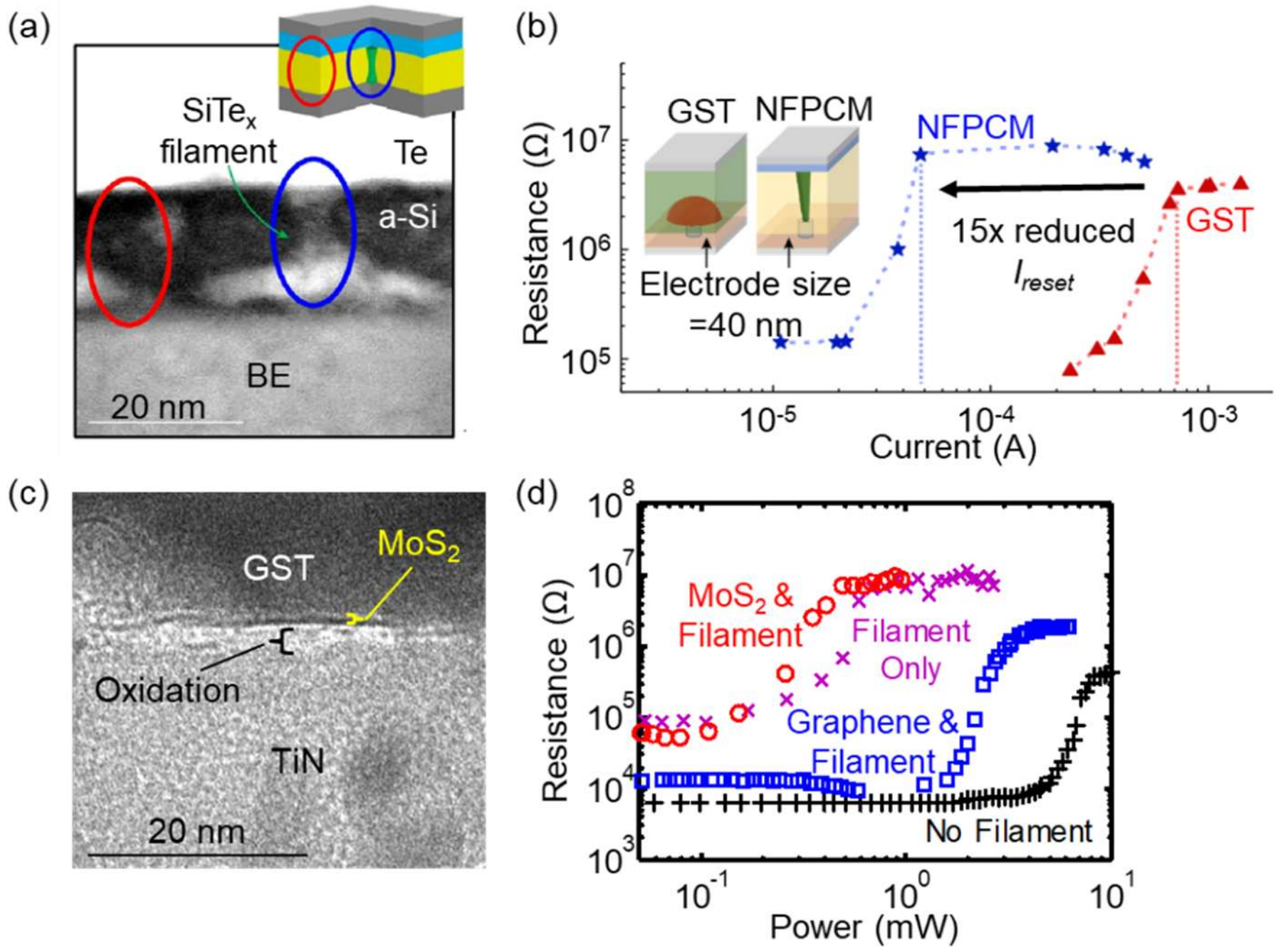

*Figure* 4 **Filamentary PCM.** (a) High-angle annular dark-field transmission electron microscope (TEM HAADF) image of the nano-filament PCM (NFPCM) in the set state, showing the Te-injected filament with ~5.5 nm diameter, and formed through the amorphous silicon (a-Si) layer *[48]*. (b) Reset transition along with the programming current of the NFPCM and the GST-based PCM showing that the NFPCM has about a 15 times lower reset current compared with the GST-based PCM with the same BE size (programming pulse width of 1 µs). The inset shows illustrations of the NFPCM and GST-based PCM fabricated for this experiment. A 10 nm a-Si is utilized for the NFPCM while the GST thickness is 40 nm. (c) TEM image of vertical PCM with oxide filament and molybdenum disulfide ($MoS_2$) inserted between GST and the metal heater *[47]*. (d) Power reduction in reset by filament formation and 2D thermal barrier. The change in read resistance after reset pulses with varying power for PCM devices with oxide filament and $MoS_2$ (red), oxide filament and graphene (blue), oxide filament only (purple), and a control device with no filament or 2D material (black) *[47]*. Figures adapted with permission from *[47], [48]*, AIP Publishing and Springer Nature.

## c) Theoretical Limits of Thickness and Diameter

Research has shown that as PCM dimensions are scaled down, the melting temperature ($T_M$) decreases, while the crystallizing temperature ($T_C$) increases [36], [54]. This defines a critical thickness below which the PCM becomes unstable in both its crystalline and amorphous phases. Nevertheless, stable crystallization has been successfully demonstrated in ultrathin films of phase change materials, such as GST (2 nm), N-doped GST (2 nm), $Ge_{15}Sb_{85}$ (1.3 nm), $Sb_2Te$ (1.5 nm), and Ag- and In-doped $Sb_2Te$ (1.5 nm) [55] as well as GeTe nanoparticles (2 nm) [56]. For GST-based PCMs, the amorphous phase becomes unstable (*i.e.*, spontaneous crystallization) when the active layer thickness is reduced below approximately 2–3 nm, as indicated by both device measurements and simulations [57], [58]. Additionally, simulations suggest that GeTe films could be scaled down to ~3.8 nm thickness (12 atomic layers); however, for thinner films, the metal-induced gap states from the two TiN electrodes overlap across the active layer, effectively narrowing the bandgap and degrading the on/off resistance ratio [59].

To reduce the volume of phase change material that must be melted during the reset operation (and thereby lower the reset energy), it is desirable to scale down both the device diameter (contact area) and the PCM thickness. However, while the PCM thickness is physically limited by material stability and performance, the device contact area is constrained by electrode contact technology. Therefore, carbon nanotubes, with diameters on the scale of single atoms, as discussed in Section 2a, approach the ultimate scaling limit in terms of contact area.

## 3. Role of Interfaces and Contacts

### a) Electrical Interfaces

PCM device scaling involves reducing all critical dimensions, including the contact diameter and the phase change material thickness. Previous studies have shown that in sub-100 nm-thick (vertical) or -long (lateral) PCM devices, the electrical resistance is dominated by the contacts [60], [61], [62], [63]. These studies employed various techniques, such as the transfer length method and cross-bridge Kelvin resistor structures to measure the resistivity, $\rho$ ($\Omega\cdot$cm) and specific contact resistivity, $\rho_c$ ($\Omega\cdot$cm$^2$) of phase change materials in both their amorphous and crystalline states.

To estimate the critical dimensions at which the bulk PCM thickness (assuming a vertical device) begins to dominate over the contact resistance, we define a characteristic thickness at which the bulk resistance equals the contact resistance: $\gamma = \rho_c/\rho$ (cm). This is an electrical equivalent of the Kapitza length often used to describe the thickness of material with the same thermal resistance as the material's interface [64], [65]. Accordingly, the proportion of electrical contact resistance relative to the total resistance can be estimated using Equation (1):

$$\frac{R_c}{R_{tot}} = \frac{1}{1 + t/\gamma} \tag{1}$$

where $R_c$ is the contact resistance in $\Omega$, $R_{tot}$ is the total device resistance in $\Omega$, and $t$ is the PCM material thickness in cm. Note that the contact resistance is measured at low bias ($V\approx0$ V) for which we assume linear contribution from both contacts, therefore a single contact contributes $R_c/2$. As shown in **Figure 5a**, when the thickness falls below ~50 nm, the contact resistance dominates the total resistance in both the crystalline and amorphous phase.

Contact resistance, $R_c$, becomes increasingly dominant as the contact diameter is scaled down (**Figure 5b**). To illustrate this, we assume that for PCM thickness below ~50 nm, $R_c$ is approximately equal to the total device resistance. Based on this assumption, the contact resistance can be estimated as $R_{tot} \approx R_C = \rho_c/A$, where $\rho_c$ is the minimum specific contact resistivity reported in the literature [60], [63] and $A$ is the contact area. Experimental measurements also show increasing resistance with decreasing contact diameter for both phases [46], [66], [67]. The measured amorphous resistance is lower than the estimated trend line, likely indicating that not all of the contact area is amorphized during measurements. A summary table of resistivity values and characteristic lengths, $\gamma$, for various phase change materials and metal contacts is provided in Supporting Information Section S3.

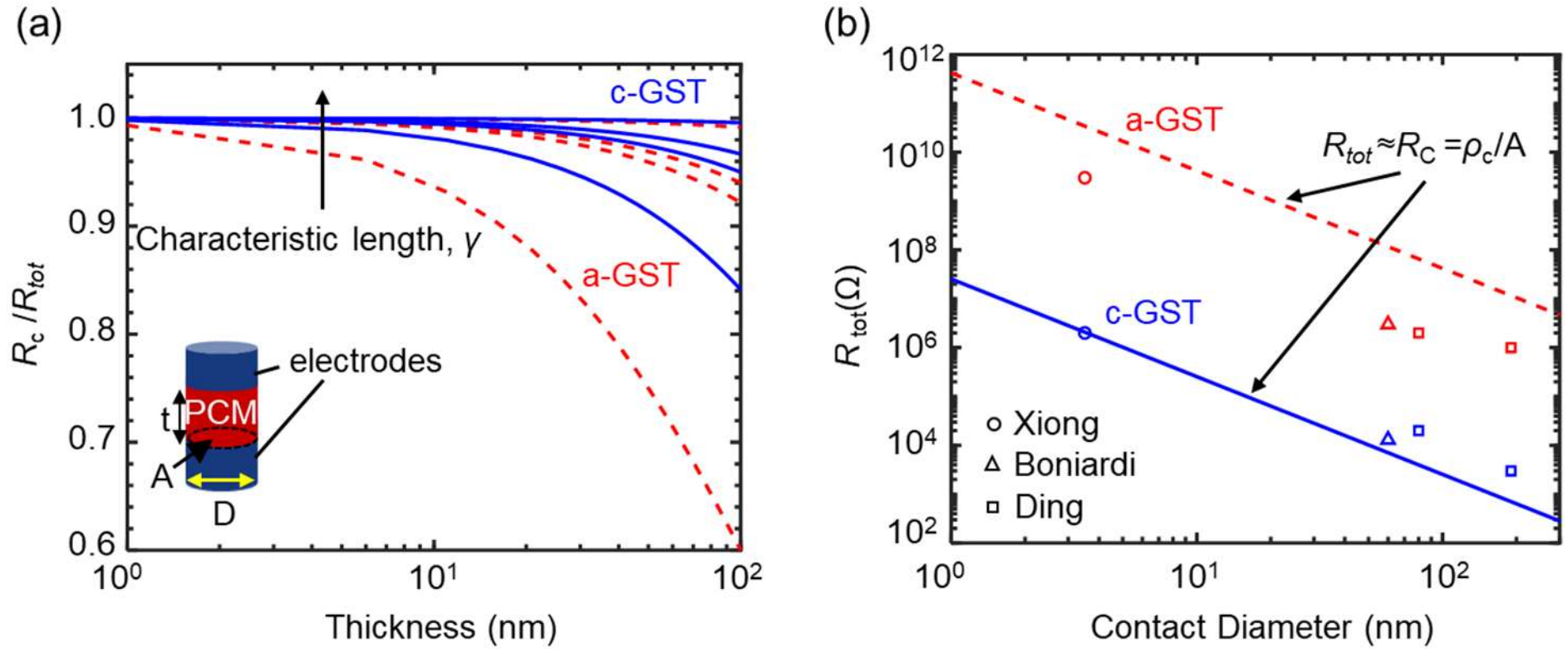

*Figure 5* **The role of contacts and interfaces for PCM scaling.** (a) Proportion of electrical contact resistance of the total resistance assuming a cylinder geometry shown in the inset based on measurements of the characteristic length, $\gamma$, for GST *[60], [61], [62], [63]*. Lowercase 'a' denotes amorphous and 'c' the crystalline phase. For thin GST $R_{tot} \approx R_c$. (b) Scaling of the total resistance, assuming it is approximately equal to the contact resistance, (amorphous and crystalline $\rho_c$ from *[60], [63]* ) with contact diameter, compared with measured resistance values from *[46], [66], [67]* . Lower measured amorphous resistance values compared to the trend line could indicate that not all of the contact area is amorphized. Results highlight the crucial role of contacts and interfaces when scaling PCM devices.

An open question still stands on the role of contacts and interfaces during switching operation. We estimate the specific contact resistivity at elevated temperatures using experimental data from room temperature up to 100 °C, and temperature-dependent *bulk* resistivity measurements up to 600 °C (**Figure 6a-b**) [60], [63], [68], [69], [70]. As expected, the specific contact resistivity decreases with temperature. Yet, the proportion of the contact resistance out of the total resistance at the elevated temperature reached during switching, has not been measured to date.

Our previous work showed that for sub-100 nm devices the non-ohmic amorphous resistance, at room temperature, can be dominated by a thermionic emission mechanism at the contacts [60]. Under this mechanism, the current follows the relation [71]:

$$I = A \cdot A^{*} \cdot T^{2} \exp\left[-q \frac{\phi_{B} - \left(\frac{qE}{4\pi\varepsilon}\right)^{0.5}}{k_{B}T}\right] \quad (2)$$

where $A$ is the contact area, $A^{*}$ is the Richardson constant, $T$ is temperature in K, $k_{b}$ is the Boltzmann constant, $\phi_{B}$ is the Schottky barrier height, $E$ is the electric field, and $\varepsilon$ is the dielectric constant of the PCM. It is likely that at high temperatures, such as those reached during switching, increased thermionic emission at the contact is less of a bottleneck for current flow, *i.e.* the contacts no longer limit the conduction. However, the exact role of the contacts during switching remains to be determined.

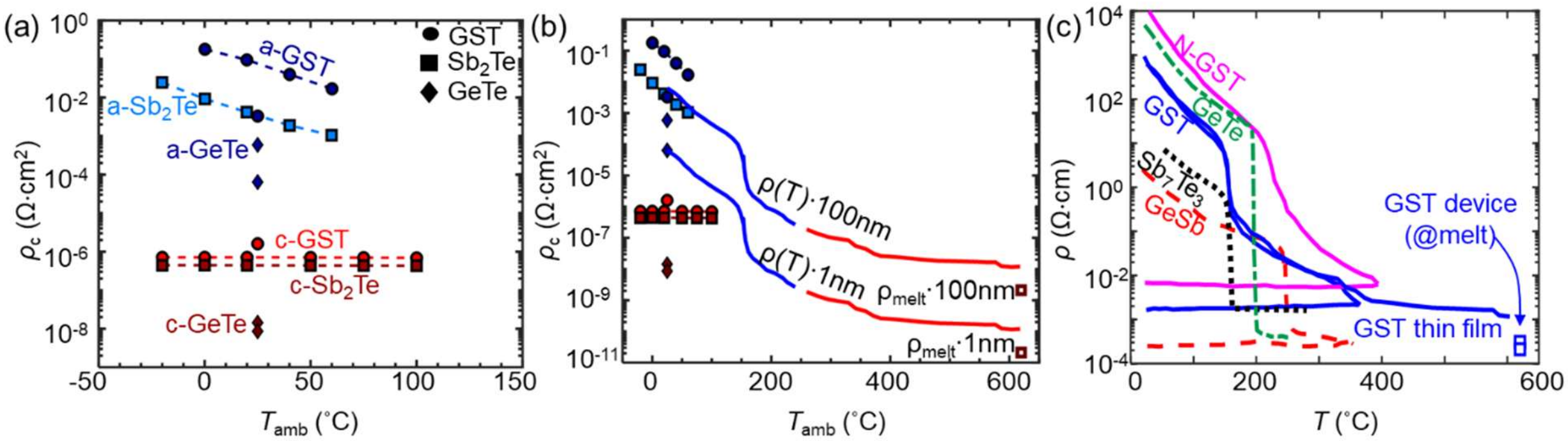

*Figure 6* **PCM temperature-dependent resistivity and contacts**. (a) Reported values from literature *[60], [63], [68]* (markers) of contact resistivity to PCM, including temperature-dependence. Lowercase 'a' denotes amorphous and 'c' the crystalline phase. (b) Data from (a) in addition to calculated resistance per area of 1 nm and 100 nm thick GST including temperature dependence up to the melting temperature, taken from thin film measurements (*[70]* solid lines). The maroon square markers represent 1 nm and 100 nm thick melted GST (resistivity extracted from bridge devices during reset *[70]*). (c) Resistivity vs. temperature of PCM materials.

## b) Thermal Characterization and Thermal Interfaces

Thermal device and material parameters play a major role in PCM switching, which is a thermally-driven process [37], [72]. Scaling PCM cells to nanometer dimensions reduces their heat capacitance and thermal time constant, which improves both energy consumption and switching speed. Nonetheless, only a limited number of studies have explored the temperature rise, electro-thermal properties, and thermoelectric effects in nanoscale PCM cells [73], [74], [75], [76]. Electro-thermal simulations quantified the impact of thermal interfaces which can reduce power consumption by ~20-30% [73], [74] for a given size, and thermoelectric effects may further reduce power consumption by ~15% of nanoscale PCM cells [75], [76]. Notably, these effects become increasingly dominant as device dimensions scale down.

Initial experimental studies relied on indirect methods for evaluating temperature [76], [77], [78]. These studies were valuable in several ways: coupling measurements with modeling to predict thin-film PCM behavior [76], highlighting the importance of cell interfaces [73], [74], and identifying both Joule heating and thermoelectric effects, including Peltier [79], Seebeck [80] and Thomson [78] within PCM cells. Despite great interest in PCM energy efficiency, direct thermal measurements of PCM devices on the nanometer-scale with nanoscale (or nanosecond) resolution are presently unavailable. However, recent work has shown temperature rise of micrometer-scale lateral GST-based devices with nanoscale resolution using an atomic force microscopy (AFM) based measurement of nanometer-scale temperature fields known as scanning Joule expansion microscopy (SJEM) [74], [81]. SJEM measurements revealed nanoscale temperature rise in such PCM test structures, and their contacts showed both Joule and thermoelectric effects [82].

Another study thermally characterized micrometer-scale lateral GST-based PCM devices using a combination of scanning thermal microscopy (SThM) and Raman thermometry [83]. This approach combined the high spatial resolution of the SThM with the material selectivity of Raman thermometry, and revealed that electrical and thermal interfaces dominate the operation of such devices with significant heating at the contacts. Similar to the SJEM results, contact heating was asymmetric, depending on current flow direction due to thermoelectric effects, highlighting that these effects must be considered when designing PCM programming pulses.

Furthermore, heat dissipation in nanoscale devices is limited by the thermal resistance of their interfaces. The thermal boundary resistance (TBR) was extracted for interfaces of both PCM-metal (*e.g.*, GST–Pt) and PCM-dielectric (*e.g.*, GST–$SiO_2$). In particular, the GST–$SiO_2$ interface had a thermal resistance equivalent to a ~50 nm thick $SiO_2$ layer (the Kapitza length), accounting for over 25% of the total device thermal resistance [83]. As device dimensions are reduced, TBR is expected to become an even larger fraction of the total thermal resistance, similar to how electrical contact resistance scales. A summary table of TBR values reported for GST with various interfaces can be found in Supporting Information section S4.

It is important to note that the thermometry experiments discussed above were conducted only in the crystalline phase, at thermal steady-state, at or slightly above room temperature. Future thermometry work should be extended to the amorphous phase, as well as to dynamic measurements over shorter transients, ideally on the nanosecond timescale, comparable to the thermal time constant. Other challenges include decoupling thermal and electrical effects, and performing measurements at actual operation temperatures (*e.g.*, near $T_M$ during the reset process). We expect thermoelectric effects to be less significant than Joule heating at elevated temperatures, because thermoelectric effects scale linearly unlike Joule heating which scales quadratically with current. However, we do not anticipate a significant change in TBR values under such conditions, and we expect the TBR to be nearly constant above room temperature, due to the relatively low Debye temperature of PCMs [84], [85].

### c) Electro-thermal and Interface Engineering in Superlattices

Enhancing thermal and electrical resistances within PCM can lead to higher energy-efficiency due to stronger heat confinement. This can be engineered *within* the active phase-change medium using superlattices (or, phase-change heterostructures), where alternating thin layers of chalcogenide materials such as $Sb_2Te_3$ or $TiTe_2$ and GeTe or $Ge_xSb_yTe_z$ are stacked, forming vdW-like interfaces between these layers (**Figure 7a-b**) [86], [87]. PCM devices based on such superlattices (SL) or heterostructures are referred to as superlattice (SL)-PCM; also, sometimes known as interfacial PCM [88] because of their interface-dominated switching mechanism.

The chalcogenide SLs can be realized at a relatively low deposition temperature of < 200 °C using sputtering. These SLs have low cross-plane thermal conductivity and very high electrical anisotropy due to the numerous parallel vdW-like interfaces, leading to strong heat confinement within the PCM device [89]. However, the interface quality depends on an optimized set of process parameters. Non-optimized deposition, *e.g.* at temperature that is too high, can lead to void formation or Te segregation [90]. In addition, the material and thermal properties of SLs are controlled by the number of interfaces and their degree of intermixing, which is found to play a crucial role in controlling the PCM device performance [86].

Leveraging the electrothermal and structural confinement in $Sb_2Te_3$/GeTe SLs, a ~ 8-10× reduction in switching current density was achieved in SL-PCM compared to control PCM with $Ge_2Sb_2Te_5$. Such SL-PCM devices with ~100 nm BE diameter have reset switching energy of ~ 6 pJ (0.3 mW reset power, with ~20 ns reset pulse) [87]. Reset pulses lead to partial amorphization near the BE of these SL-PCM devices during the low-to-high resistance state transition. However, the SL layers above and surrounding remain crystalline, which act as a template to reconstruct the vdW interfaces within the SL during the set transition. Thus, the heating efficiency in SL-PCM is maintained upon repeated switching [86], [87], [91]. Recently reported endothermic melting transition with ∼240 °C lower temperature and ∼8× lower enthalpy in such SL films compared to melting of GST, further provide key thermodynamic insights into the low-power switching of SL-based PCM [92]. Such thermally-induced phase change in SL-PCM technology is further explained and supported by the scalability of switching power with BE diameter, an important feature for ultra-scaled devices.

To this end, the optimization of the number of SL interfaces is important to achieve lower switching current and power in SL devices. The switching current has been shown to decrease with increasing number of interfaces, while a higher degree of intermixing and imperfections within the layers increase the switching current in an SL-PCM device [93], [94]. By controlling the number of SL interfaces and preserving interface quality, low switching current densities (≈ 3–4 MA/cm²) were demonstrated simultaneously in the same SL-PCM device based on an $Sb_2Te_3$/GST superlattice (2/1.8 nm/nm), with bottom electrode diameters down to 110 nm (**Figure 7c**) [93], [95].

A remaining question is whether SL-based PCMs can maintain advantages as they are reduced in size. Using a combination of phase-change material SLs and nanocomposites (based on $Ge_4Sb_6Te_7$ [96]), a switching current of ~ 85 µA at ~ 0.7 V switching voltage was shown in PCM devices with the smallest dimensions to date (~ 40 nm bottom electrode) for a SL technology [86]. These SLs consisted of 15 periods of alternating layers of $Sb_2Te_3$ (~ 2 nm) and $Ge_4Sb_6Te_7$ (~ 2 nm), sputtered at 180 °C. The efficient switching is enabled by strong heat confinement within the SL materials and the nanoscale device dimensions. The scalability of reset current with BE diameter is maintained in these SL-PCM devices even at nanoscale dimensions (from ~ 80 nm down to ~ 40 nm, as shown in **Figure 7d**), which promises further reduction of switching energy by downscaling or confining such SL-PCM [86].

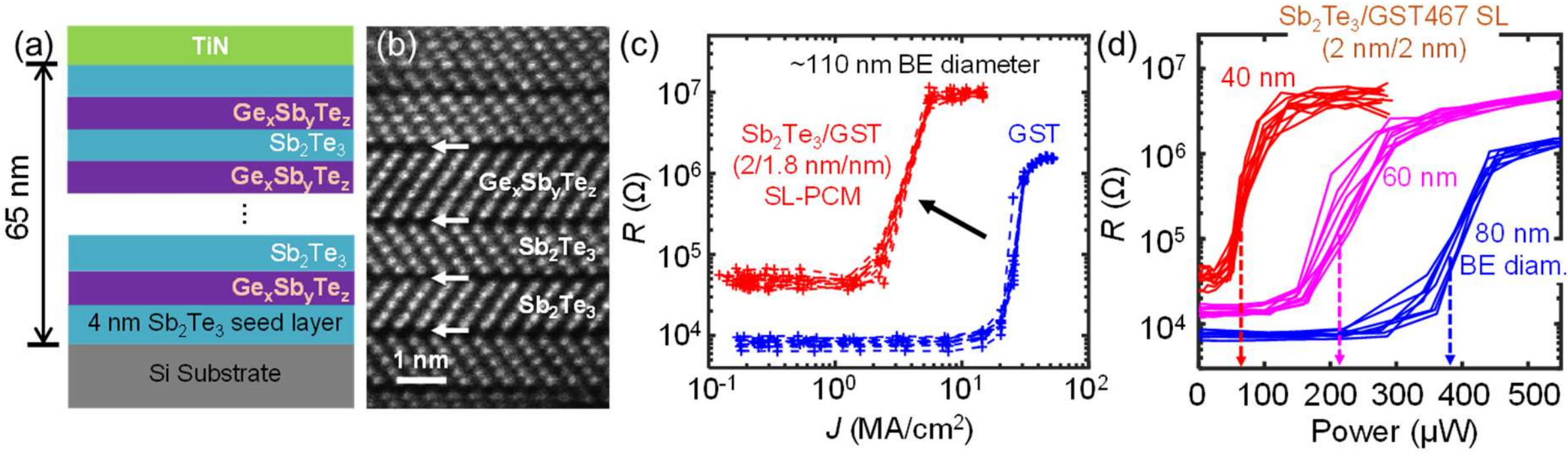

*Figure 7* **Superlattice** (a) Schematic of superlattice (SL) stack with alternating $Sb_2Te_3$ and $Ge_xSb_yTe_z$ (~65 nm total thickness) including the ~4 nm $Sb_2Te_3$ seed *[93]*. (b) High angle annular dark field scanning transmission electron microscopy (STEM) cross-sections of a 2/1.8 nm/nm $Sb_2Te_3$/ $Ge_xSb_yTe_z$ SL showing sharp interfaces and vdW-like gaps parallel to the substrate *[93]*. (c) Resistance ($R$) vs. current density ($J$) showing a significant improvement of reset current density ($J_{reset}$) in $Sb_2Te_3$/ $Ge_xSb_yTe_z$ SL-PCM vs. GST PCM (here both devices have a BE diameter of ~110 nm) *[93]*. Adapted from A. Khan *et al.*, Nano Letters, 2022, with permission from ACS Publications. (d) Measured DC read resistance ($R$) vs. power ($P$) for $Sb_2Te_3$/$Ge_4Sb_6Te_7$ SL-PCM devices with varying BE diameter (from 80 nm down to 40 nm). 10 different cycles are shown for each device in both figures. Dashed colored arrows indicate the reset power for different BE diameter SL-PCM devices *[86]*.

## 4. PCM Heat Confinement

### a) Adiabatic Energy Limit

The fundamental limit of reset energy, assuming no energy dissipation to the environment beyond the active PCM material, is defined as the reset adiabatic energy limit. From a thermodynamic standpoint, the energy density in this limit can be estimated as:

$$E_{min} = C_s \Delta T_M + H \quad (3)$$

where $C_s$ is the specific heat, $\Delta T_M$ is the temperature rise for melting, and $H$ is the latent heat of melting [12]. For GST $C_s \approx 1.34$ J·cm$^{-3}$K$^{-1}$, $\Delta T_M \approx 600$ K, and $H \approx 90$ J·cm$^{-3}$ resulting in a minimal energy of ≈0.9 aJ/nm$^3$ [97], [98] (Supporting Information Section S5). This minimal energy is the energy per unit volume needed to reset a PCM cell that is perfectly insulated thermally, *i.e.* adiabatically, from its environment. This condition corresponds to an infinite thermal boundary resistance (TBR → ∞) at the PCM interface with metal electrodes and surrounding dielectrics layers.

The minimal adiabatic energy, obtained by multiplying the minimal energy density ($E_{min}$) by the PCM volume, can be estimated using two geometries: a sphere with varying diameter, or a disk with fixed thickness (10 nm) and varying diameter, as illustrated in **Figure 8**. While the spherical model is often used for its simplicity, the disk model generally aligns more closely with experimental data. A comparative table of minimum melt energies for different phase change materials is provided in Supporting Information Section S5.

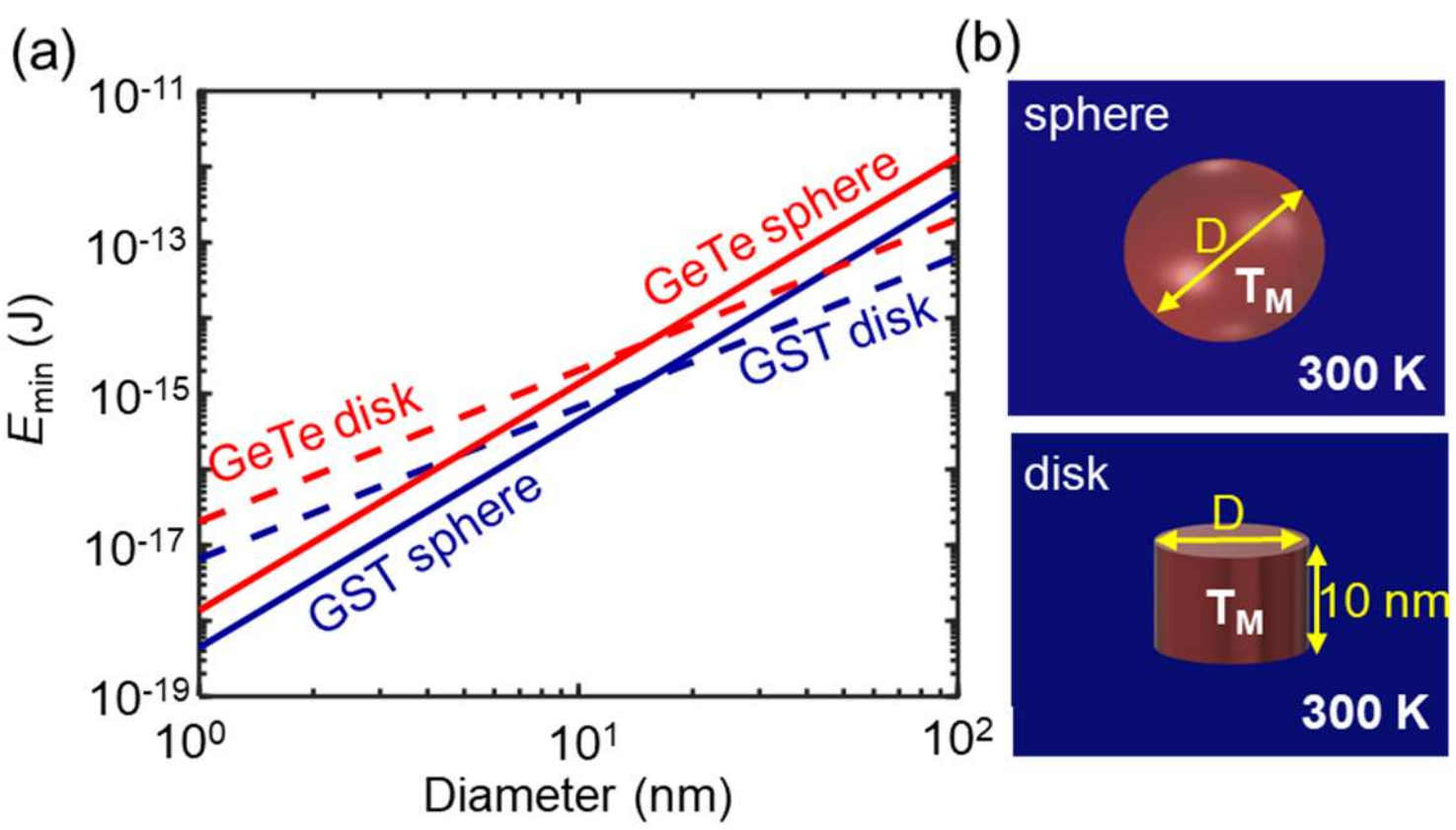

*Figure 8* **Model for estimating the adiabatic limit.** (a) Calculated minimum reset energy vs. PCM diameter, where an adiabatically isolated PCM volume is heated from room temperature to its melting temperature. The calculated energy includes the latent heating of melting. Material properties used are shown in Supporting Information Section S5. (b) Two geometries are shown: sphere and disk with a fixed thickness of 10 nm, both with varying diameter *D*. The GST disk is used later in **Figure 14** & **Figure 15**. The PCM volume is assumed to reach melting temperature $T_M$, while the ambient remains at 300 K.

## b) Improving Energy-Efficiency with Heat Confinement

The adiabatic reset energy for devices with diameters of just a few nanometers is on the attojoule scale. However, practical PCM devices are not adiabatic because they unavoidably lose heat to their surroundings, through a finite TBR. To improve the energy efficiency of PCM devices towards their fundamental limits, it is essential to engineer them for maximum heat confinement. While phase transitions in PCM are typically triggered electrically, the transformation itself is ultimately thermally driven. Unfortunately, in most device architectures, a significant portion of the thermal energy is lost to heating the surrounding materials rather than being confined to the active PCM volume. During heating with a certain pulse width, denoted $\tau_{pw}$, the thermal diffusion length into the surrounding materials is $(\alpha\tau_{pw})^{1/2}$ which is about 30 nm into $SiO_2$ for a 1 ns heating pulse, given that its thermal diffusivity $\alpha \approx 10^{-6}$ m$^2$s$^{-1}$. When PCM cells are at nanoscale dimensions, or when heating pulses are longer than a few ns, *i.e.,* longer than the thermal time constant, $\tau_{th}$, the volume of heated surrounding material becomes significant [99]. Note that in practical PCM devices, the active region cannot be fully thermally isolated, and cooling occurs primarily through the electrodes, with rapid quenching during reset and slower cooling during set. For small active volumes (sub-100 nm), this is not a limiting factor since $\tau_{th}$ is sufficiently short to allow effective quenching.

There are two main approaches to reduce energy loss to the environment surrounding the PCM, *i.e.*, to improve heat confinement: (1) increasing the thermal boundary resistance (TBR) between the PCM and its surroundings, and (2) shortening the heating pulse width, thereby limiting the heat diffusion length. In the ideal limits of TBR $\rightarrow \infty$ or pulse width $\tau_{pw} \rightarrow 0$, the system approaches adiabatic heating, as validated by finite element simulation (**Figure 9**). For the spherical model, heat dissipation can also be calculated analytically, as shown in Supporting Information Section S6. **Figure 10**, based on these simulations, further demonstrates how increasing TBR and reducing pulse width significantly improves energy efficiency. The discussion focusses on the reset process which typically consumes higher programming energy and is not limited by crystallization time. Set energy is discussed later in Section 5a.

In addition to thermal interface properties and pulse length, cell dimensions also influence heating efficiency. As shown in **Figure 11**, the thermal healing length, the characteristic distance over which a temperature gradient persists (~30 nm in typical PCM mushroom structures), is larger than the small device with 10 nm BE diameter, but represents only ~30% of the large device with 100 nm BE diameter. For the same pulse width, the fractional heat spreading is greater in the smaller device, resulting in lower efficiency. In other words, smaller devices also require shorter reset heating pulses in order to be more energy-efficient.

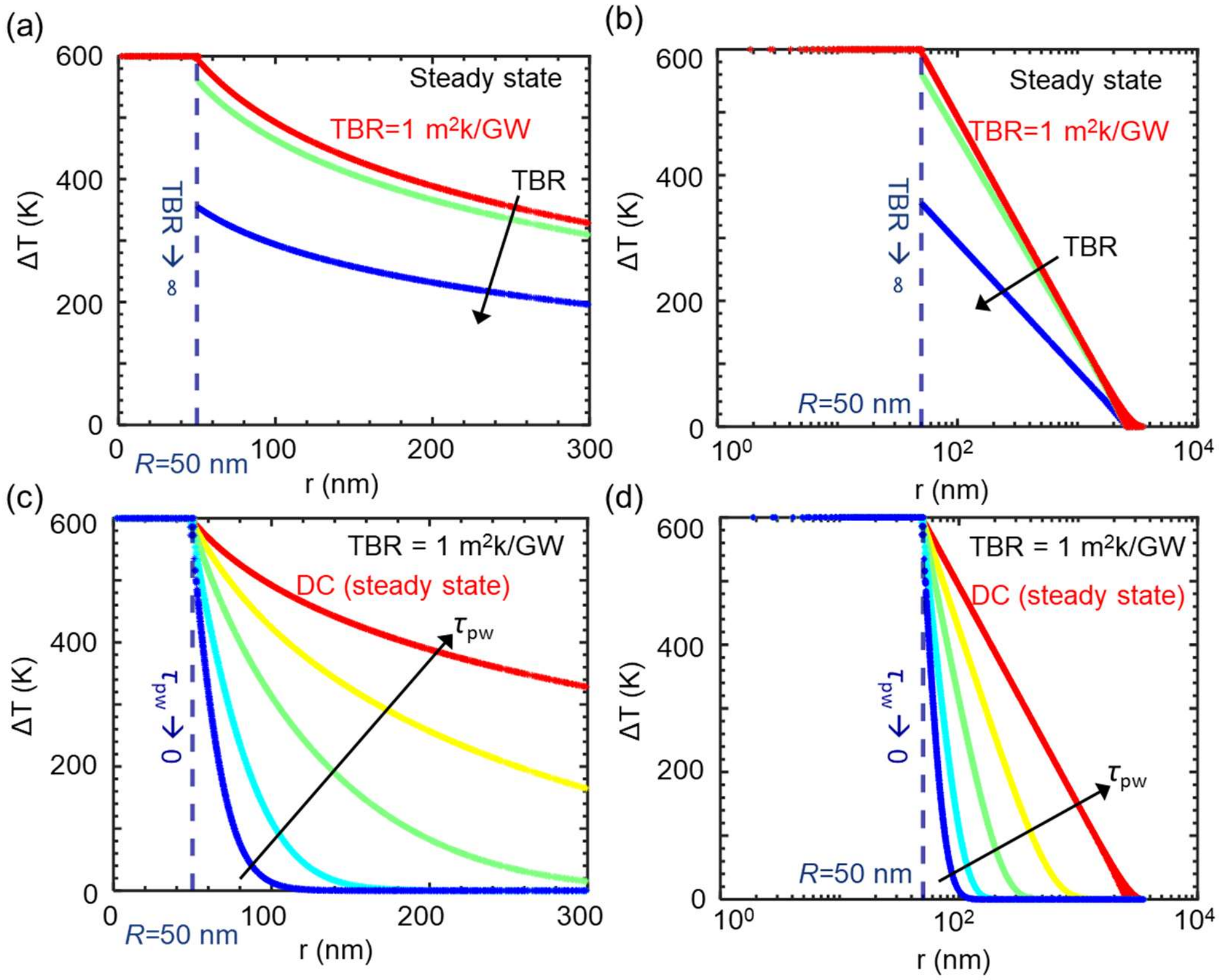

*Figure 9* **Finite element simulation of heat dissipation.** Temperature rise as a function of distance from the origin in a spherical PCM device with $D$ = 100 nm (radius $R = D/2$), for varying TBRs (= 1, 10 and 100 m$^2$K/GW) assuming steady state conditions: (a) linear scale (b) semilogarithmic scale of horizontal axis. Temperature rise for varying pulse widths (including DC and PW of 100, 10, 1, 0.1 ns) assuming TBR = 1 m$^2$k/GW: (c) linear scale (d) semilogarithmic scale of horizontal axis. The area under each curve (*i.e.*, the integral) is proportional to the energy dissipation. Vertical dashed lines mark the adiabatic limit, with TBR → ∞ or $\tau_{pw} \to 0$.

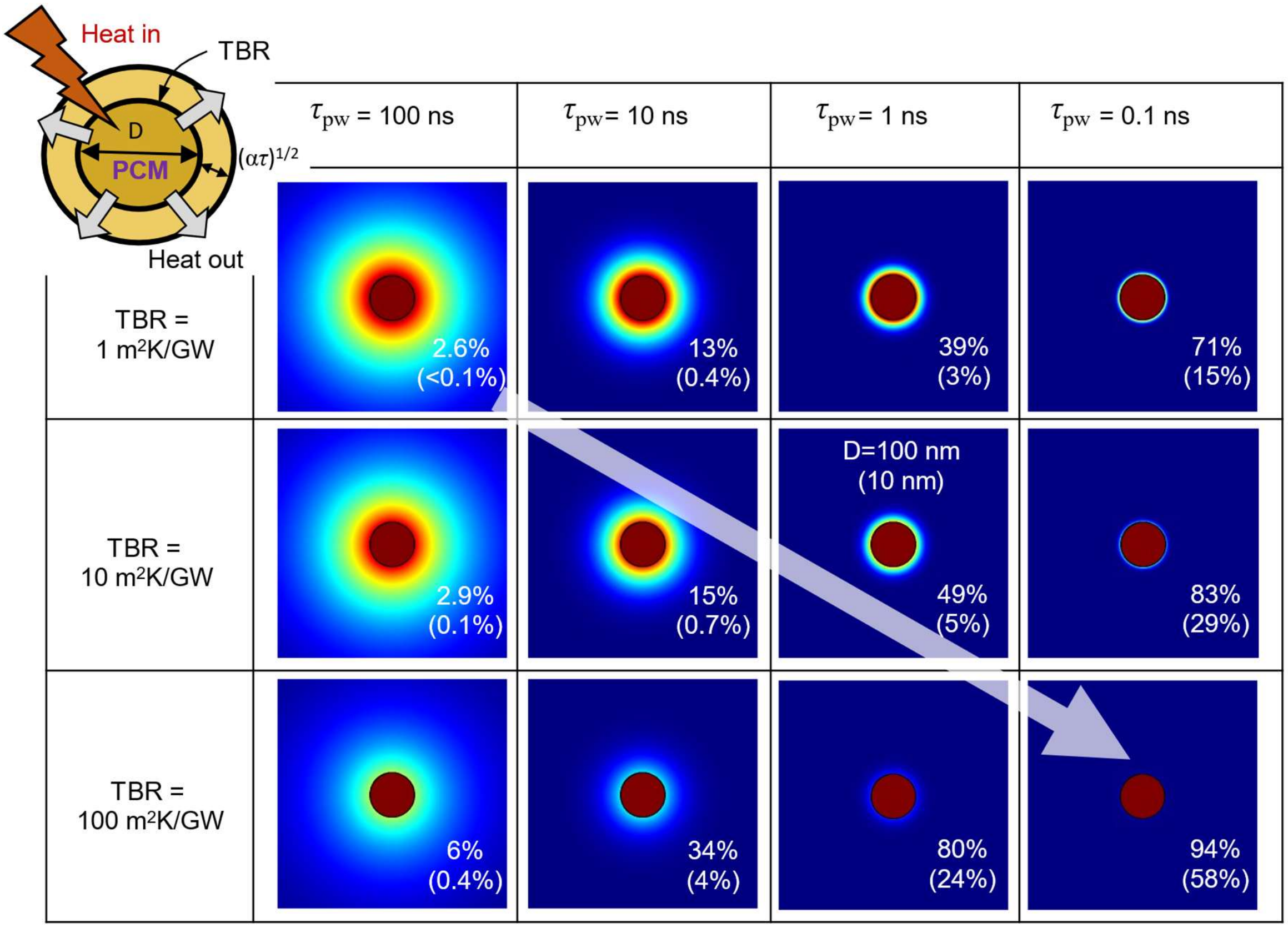

*Figure 10* **PCM energy efficiency explained by simplified model.** Comparison of heating profiles and energy efficiency (defined as the ratio of PCM melting energy to total input energy) using a simplified spherical model of PCM embedded in $SiO_2$. Values are shown for spherical diameter $D$ = 100 nm (with corresponding $D$ = 10 nm values in parentheses), for varying thermal boundary resistance (TBR, horizontal rows) and programming pulse width ($\tau_{pw}$, vertical columns). Block arrow shows trend of heat confinement and energy efficiency improving with increasing TBR and decreasing $\tau_{pw}$. Note that following a single column along the r-axis corresponds to **Figures 9a–b**, while following a single row corresponds to **Figures 9c–d**.

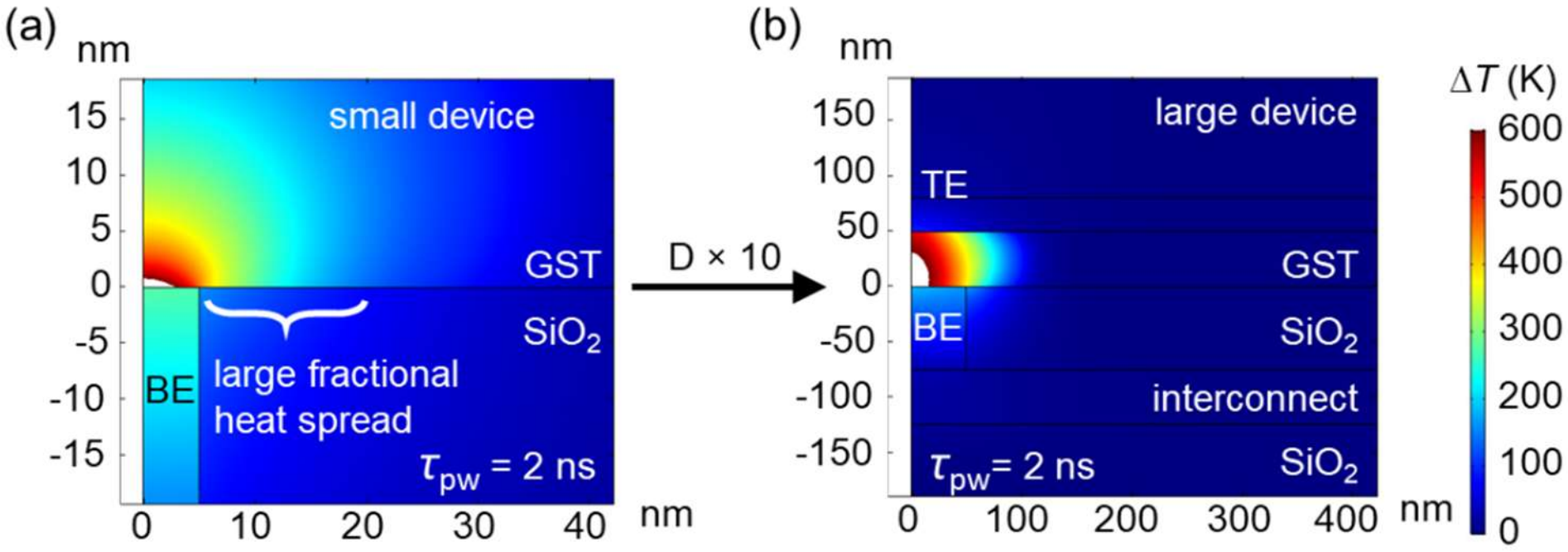

*Figure 11* **Heating efficiency dependence on cell size in a realistic structure.** Finite element simulation of a typical 50-nm-thick GST mushroom cell with BE diameter of (a) 10 nm and (b) 100 nm. Note the different axis ranges. The left vertical axis is the axis of symmetry of the cylindrical structure. The thermal healing length (~30 nm) is larger than the BE diameter of the small device and ~30% of the size of the large device. For the same pulse width, the fractional heat spreading of the small (10 nm) device is much larger and hence its efficiency is lower.

### c) Heat Confinement by Thermal Resistance

Much research has been devoted to find ways to create higher thermal resistance, consequently reducing dissipated heat and energy loss in PCM. Work by Matsui *et al.* [100] shows that the dissipated cell power during switching is significantly reduced by inserting a very thin $Ta_2O_5$ film between the GST and the bottom W electrode plug. The $Ta_2O_5$ interfacial layer functions as a thermal insulator, preventing heat generated in the programming volume of the GST from diffusing to the W plug, which has high thermal conductivity [100]. Furthermore, Fong *et al.* [101] utilized a sequence of $SiO_2/Al_2O_3$ dual-layer stacks (DLS) for better heat confinement, due to lower thermal conductivity of the multilayer insulator compared to $SiO_2$ alone (**Figure 12a**). The dashed confined devices using the DLS thermal insulator require ~60% less reset energy than devices with $SiO_2$, showing low reset energies of 18.25 $\pm$ 15.8 pJ for 10,000 $nm^2$ contact area, and reset current densities of 0.94 $\pm$ 0.51 $MA/cm^2$. This work also showed reduced reset energy though scaling as discussed previously in Section2.

Thermal insulation was also implemented with atomically thin interfacial layers, such as 2D $MoS_2$, presented previously in Section 2b (**Figure 4c-d**) [47]. The $MoS_2$ interfacial layer performs better than a graphene interfacial layer because $MoS_2$ has lower thermal conductivity and higher TBR. Another study [75] used a thin $Bi_2Te_3$ interfacial layer to implement thermoelectric heating, achieving a ~2× reduction of reset current density and power vs. the control GST-based devices.

As discussed previously in Section 3c, superlattices employ the concept of heat confinement to enhance energy efficiency. Heat confinement in SL-PCM devices can be further improved by using a pore-type structure on a thermally insulating flexible substrate (**Figure 12b-c**), where the SL phase-change material limits heat loss into the metal electrodes, the very low thermal conductivity of the flexible substrate blocks heat loss from the BE, and the pore-like geometry provides Joule heat confinement [102]. By directly fabricating the SL-PCM devices with a pore-like geometry (~600 nm BE diameter) on a flexible polyimide substrate, a switching current density of 0.1 $MA/cm^2$ was achieved, the lowest recorded to-date in PCM technology (**Figure 12d**) [102]. Notably, the flexible SL devices were fabricated without any layer transfers and at a maximum process temperature of 200 °C.

Recently, low switching current in mushroom-cell PCM devices has also been reported using GST grown on a highly-oriented 2 nm thin $TiTe_2$ seed layer [103]. This study also leveraged heat confinement from the vdW interfaces enabled by highly oriented homostructure GST. Such homostructure devices show a similar reset current density as of a SL PCM device with alternating films of $Sb_2Te_3$ and GST (5 nm/5 nm), with the same BE size [86]. While such $TiTe_2$/GST bilayer films are less prone to deposition and process-induced variability [103], SL hetero-structures with thinner layers (*i.e.*, higher interface density) have shown advantage of lower switching current density and voltage at the same device dimension [86].

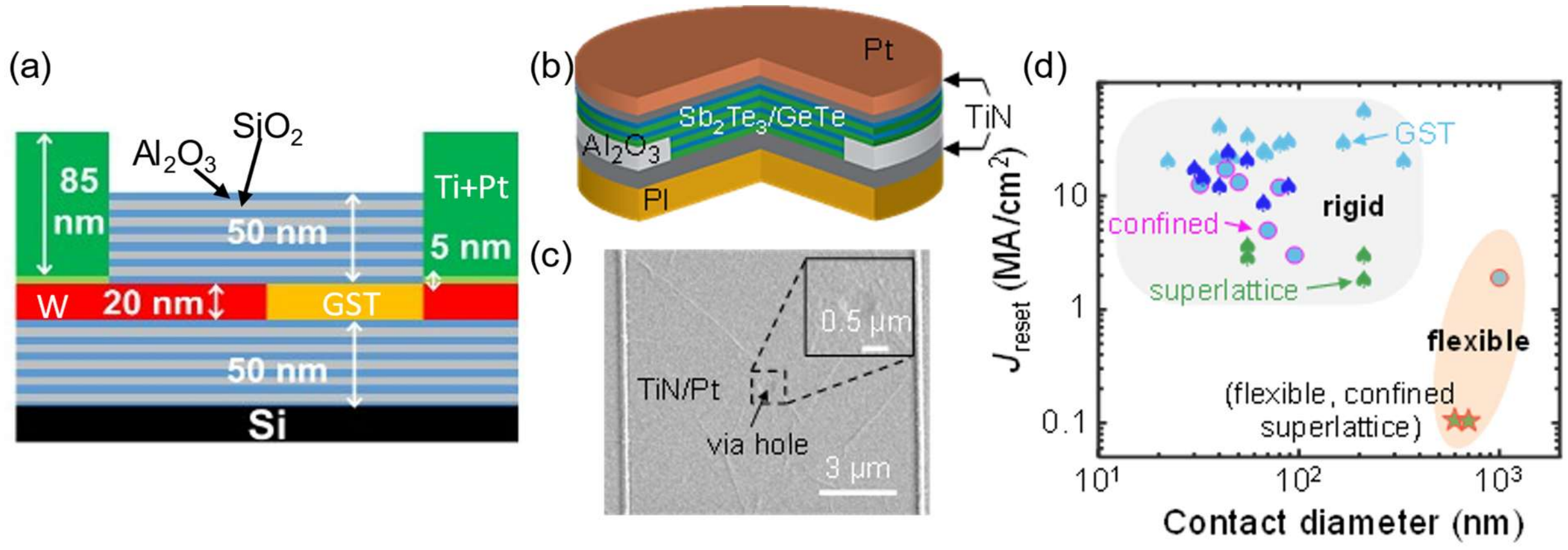

*Figure 12* **Heat confinement by increased thermal resistance**. (a) Schematic side profile view of dash-confined lateral GST device with $SiO_2/Al_2O_3$ dual-layer stack (DLS) *[101]*. Flexible SL-PCM *[102]* (b) schematic device cross-section of $Sb_2Te_3$/GeTe SL-PCM on a flexible polyimide substrate, and (c) Top view scanning electron microscope image of such a flexible SL-PCM device with ~600 nm via hole (inset shows zoomed view). (d) Benchmarking of reset current density ($J_{reset}$) reveals around two orders of magnitude lower $J_{reset}$ in flexible, confined SL-PCM compared to conventional PCM on rigid silicon substrates and one order of magnitude lower than other flexible PCM. Adapted from A. Khan *et al.*, Science, 2021, with permission from AAAS.

## d) Heat Confinement by Reducing Pulse Width

Heat confinement and energy efficiency in PCM devices can be significantly improved by using sub-nanosecond reset pulses. Such short pulses limit the heat diffusion time, thereby minimizing thermal dissipation to the surrounding environment. Sub-ns pulses can be readily applied and measured using on-chip circuitry; however, when accessing the device via contact pads in a probe station, a specialized high-speed setup is required. This includes radio frequency (RF) probes to minimize parasitic capacitance and enable impedance matching, thereby suppressing signal reflections, as shown in **Figure 13a**. To ensure a resistance-capacitance (RC) delay shorter than ~1 ns, devices must be sufficiently small, and the contact pads must be capacitively decoupled from the substrate, especially if it is conductive.

The concept of improved energy efficiency through sub-ns pulse application has been experimentally demonstrated in confined PCM cells with a bottom electrode (BE) diameter of 50-200 nm (**Figure 13b**) [99]. A key finding of this work is that the switching power remains constant for pulse widths longer than the PCM thermal time constant, $\tau_{th}$, which is approximately 1 ns for a device with a 50 nm BE diameter. The thermal time constant is the characteristic time scale over which the PCM temperature changes in response to variations in input power. It is given by the product of thermal capacitance (which depends on the heated volume) and thermal resistance. For PCM, the relevant volume defining $\tau_{th}$ is the critical volume required to induce a measurable change in electrical resistance through a phase transition. Energy consumption is reduced as pulse width is decreased down to ~ $\tau_{th}$, as shown in **Figure 13c** [99], owing to improved thermal confinement achieved by limited heat diffusion time. This programming scheme reduces the reset energy density to below 0.1 fJ/nm$^2$ with a pulse width of 0.3 ns, more than an order of magnitude lower than state-of-the-art PCM devices with the same BE diameter [67], [99].

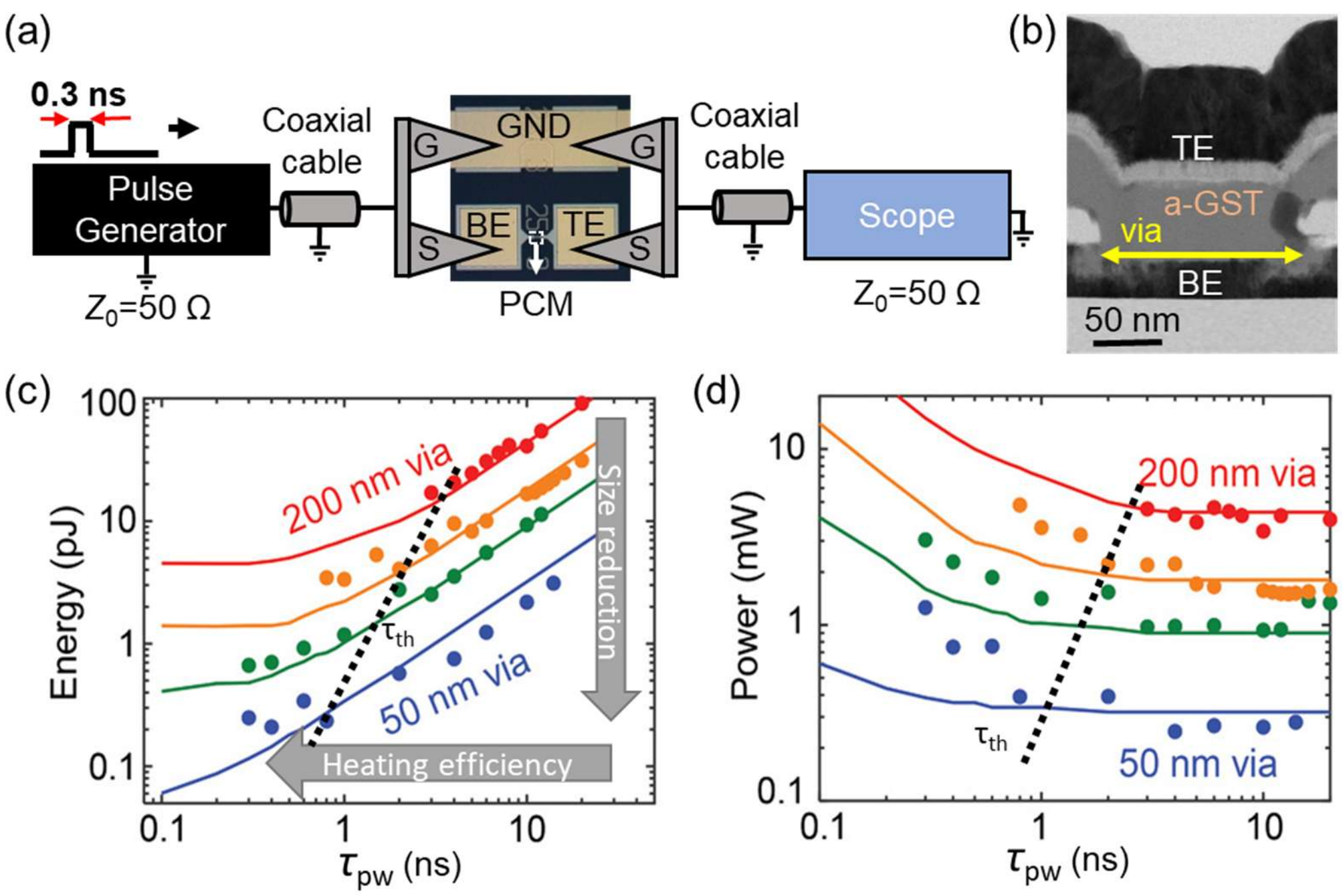

*Figure 13* **Sub-nanosecond programming pulses improve energy efficiency**. (a) Schematic of high-speed measurement setup. A pulse generator with output resistance of 50 Ω is connected by coaxial cable to an RF probe with the signal (S) to the bottom electrode (BE) pad of the PCM device and ground (G) to a local ground plane, shown in the optical image of the device. A fast scope is connected by RF probes (S) to the TE of the device. The RF probes are needed to minimize parasitic capacitance, and the 50 Ω impedance matching suppresses signal reflection *[99]*. (b) Bright field scanning transmission electron microscope (BT-STEM) cross-section of confined device in the amorphous phase *[60]*. Log-log measured (markers) and simulated (lines) (c) reset energy and (d) reset power vs. pulse width ($\tau_{pw}$) for devices with different via size (nominal ≈ 50, 75, 100, 200 nm). The thermal time constant, which is size-dependent, is marked with a black-dashed line, as a guide to the eye *[99]*. Adapted from K. Stern *et al.*, Adv. Electron. Mater., 2021, with permission from John Wiley and Sons.

The thermal time constant $\tau_{th}$ is a key parameter for understanding the trade-off between reset energy and power. For switching pulses longer than $\tau_{th}$ the reset power remains approximately constant, allowing energy consumption to scale linearly with pulse width ($\tau_{pw}$). However, for faster switching, when using pulse widths shorter than $\tau_{th}$ a higher reset power is required to fully melt the GST, as shown in **Figure 13d**. In this regime, we observe only a slight reduction in reset energy with decreasing pulse widths, much less pronounced than the energy savings achieved when reducing pulse widths above $\tau_{th}$. Therefore, for pulse widths shorter than the thermal time constant, each application must weigh its priorities, whether to optimize for power, energy efficiency, or speed.

In the sub-ns switching regime, the PCM reset energy approaches the charging energy of interconnect lines in a crossbar array (dashed line in **Figure 2a**), serving as a practical lower-limit for all NVM and neuromorphic devices. This highlights that with appropriate modifications such as device scaling, enhanced thermal resistance, and reduced pulse widths, the reset energy is not necessarily the primary bottleneck, and therefore set energy is discussed in the next section.

## 5. Ultimate Energy and Scaling Limits of PCM

### a) Energy of Set Process

Until now, we have primarily focused on the reset energy of PCM, that is, the energy required to amorphize the material through a melt–quench process, since this operation requires higher temperatures and therefore higher input power. However, the energy required for the set process, which recrystallizes the phase-change material, is not negligible. A simple back-of-the-envelope calculation suggests that the set power is approximately three times lower than the reset power, based on the required temperature rise, $\Delta T_{\mathrm{M}} \approx 3\Delta T_{\mathrm{C}}$ [36] implying $P_{\mathrm{reset}} \approx 3P_{\mathrm{set}}$ for PCM devices. However, the set operation is fundamentally limited by crystallization dynamics, involving nucleation and growth, with typical set times exceeding 50 ns. In contrast, reset times can approach or fall below 1 ns. As a result, the set time remains more than an order of magnitude longer, making the set energy ($E = P{\cdot}\tau_{\mathrm{pw}}$) a significant contributor to total energy consumption. Therefore, to further improve PCM energy efficiency, substantial improvements in set speed are essential.

Previous studies have explored several approaches to shorten the crystallization time in PCM devices [30], [104], [105], [106], [107], [108], [109], [110], [111], [112]. Reducing the device size has been shown to reduce set time [30], [104], [105], with reported values down to ~2 ns for cells with ~20 nm diameter. In comparison, reset pulses for the same device size can be as short as ~0.4 ns, nearly an order of magnitude faster. Thus, while scaling improves write energy for both operations, set energy is still likely to remain higher than reset energy. An alternative approach involves applying a priming electric field to reduce the set time [106], [107]; however, this technique is less suitable for conventional operation because it requires higher idle power and, consequently, higher energy consumption.

Another strategy is to optimize the phase-change material composition relative to standard GST. For example, scandium-doped GST [108], reduced the set time from 33 ns (at 2.1 V) to 6 ns (at 2.5 V) for a 80 nm contact diameter. Work using scandium–antimony–telluride [111] reported even faster switching speeds of 700 ps. However, scandium is not commonly used in standard semiconductor processes. Boniardi *et al.* [109] demonstrated a reduction in set time to 80–100 ns (from 180–200 ns for standard GST) by increasing the antimony (Sb) concentration. This improvement came at the cost of higher switching current and a narrower programming window. Other materials, such as GeTe [110] achieved for a 60-nm diameter device set times as low as 16 ns, and even 1 ns for devices with lower initial resistance.

Recent work [112] achieved a set time of 650 ps in laterally-confined GST devices using a $TiO_2$ dielectric layer. Zhao *et al.* [112] attributed this improvement to the octahedral atomic configuration of $TiO_2$, which may provide nucleation interfaces that enhance crystal growth of GST. However, the high stochasticity of the reset state and low on/off ratio in that study raise questions about whether the observed switching is due to a phase change or resistive switching in the $TiO_2$ itself. Further studies on crystallization dynamics [113] may aid in developing a deeper understanding of the set process and in reducing energy consumption through improved material and device engineering.

## b) Summary of Set and Reset Energy

Before concluding, we summarize the energy consumption trends reported across the various studies discussed in this review. Energy consumption is reduced primarily by minimizing the volume of phase-change material involved in switching (*i.e.*, via scaling) and through improved heat confinement. Enhanced heat confinement reduces the energy lost to the surrounding environment, which can be achieved by increasing the device thermal resistance and applying shorter pulse widths. We can estimate the energy consumption during reset and set as:

$$E_{reset/set} = \int_0^{\tau_{reset/set}} I_{reset/set}(t) \cdot V_{reset/set}(t) dt \approx I \cdot V \cdot \tau_{pw} \tag{4}$$

where $I$ and $V$ represent the transient current and voltage during the reset or set operation, and $\tau_{pw}$ is the pulse width. Probing the transient current during operation is essential, as the PCM resistance in the molten state during reset is lower than its resistance in the crystalline state. Consequently, the reset current cannot be estimated using only the applied voltage and the crystalline-state resistance. For this reason, we only report values from studies that provide detailed measurements of current, voltage, and pulse width *during* switching. Importantly, we use the actual voltage drop across the PCM device, not the total applied voltage, as the two can differ significantly depending on the measurement setup and device structure. **Figure 14** summarizes the reset and set energies reported in this manuscript, categorized by material composition and device geometry. The fundamental lower limit, shown in **Figure 14** and **Figure 15** corresponds to the adiabatically isolated disk case discussed earlier.

Low reset energy consumption is observed in **Figure 14b** for SL devices (green right-pointing triangles), and for confined GST with sub-ns pulses (blue square), demonstrating the effectiveness of the heat confinement. In SL devices, this is attributed to the numerous parallel vdW-like interfaces, while in confined GST, it results from short pulses that limit heat diffusion. Additionally, we consider the critical dimension for each architecture (diameters shown in **Figure 14** and **Figure 15**) to evaluate the scaling behavior of switching energy, as depicted in **Figure 1c** and Supporting Information **Figure S1**. It is evident that the energy consumption decreases by reducing the critical dimension, with CNT devices (blue stars) showing the lowest measured values for PCM devices (~tens of fJ). **Figure 14** and **Figure 15** do not include data for filamentary devices, as their precise critical dimensions are not well known.

Currently, the set energy exhibits a significantly larger gap between practical PCM devices and the adiabatic limit than the reset energy, as highlighted in **Figure 14a**. This arises because, in practice, set and reset energies are often comparable, and in some cases the set energy is even higher, despite the adiabatic limit for set being approximately three times lower than that for reset. As discussed earlier, the long set time is the primary contributor to this disparity between practical PCM performance and the physical limit.

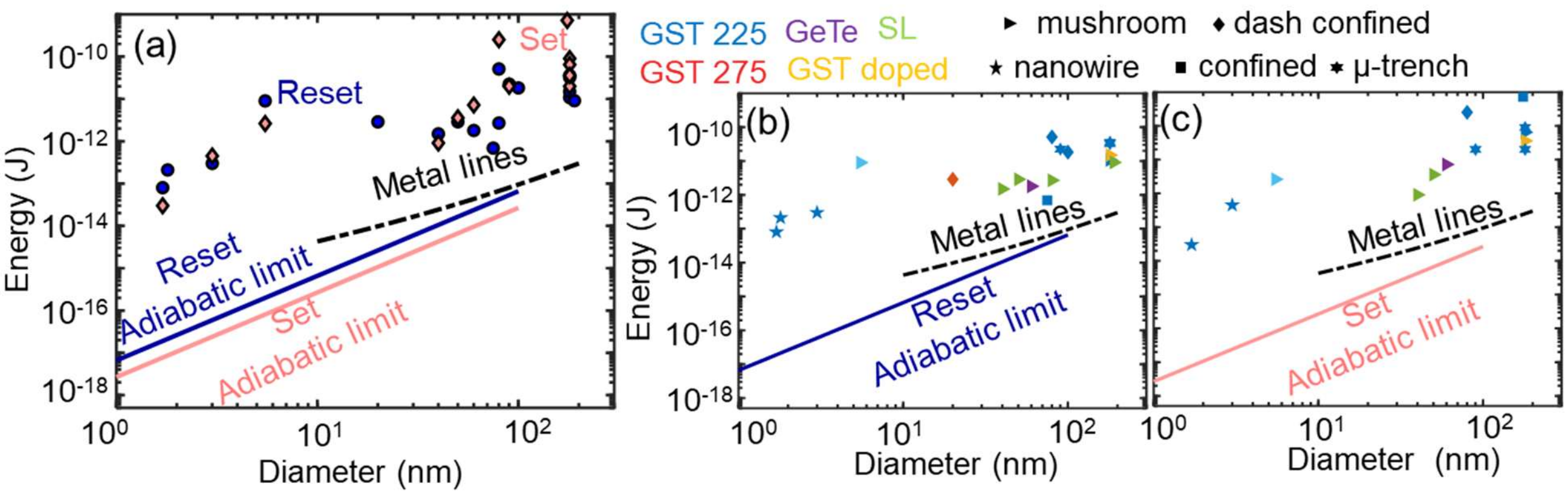

*Figure 14* **PCM energy scaling.** Energy vs. diameter during programming for (a) reset and set (b) only reset and (c) only set. Markers represent experimental data points from literature as shown in Supporting Information Section S7, the different geometries and the relevant effective contact diameter are shown in **Figure 1c** and Supporting **Figure S1**. Lines are calculations of the adiabatic limit based on the disk model for GST described in the text. The adiabatic set energy is calculated using the same specific heat ($C_s$) and latent heat ($H$) as used for the adiabatic reset energy, just with the crystallization temperature ($\Delta T_C \approx 200$ $K$). Dashed black lines represent metal line's energy to charge interconnects of a single bit in 1k×1k array.

To further investigate the ultimate reset energy limit of PCM, we provide several estimations of the energy required to reach the reset temperature in PCM cells, varying the $\tau_{pw}$ and TBR, as shown in **Figure 15a-b**. We present finite element simulation results for a simplified spherical GST cell surrounded by $SiO_2$, which, despite its simplicity, captures the main physical trends and yields the correct order of magnitude. An analytical calculation, shown for the case of TBR = 0 $m^2KGW^{-1}$, is provided in Supporting Information Section S6. **Figure 15a** shows simulated reset energy vs. diameter for $\tau_{pw}$ ranging from 0.1 to 100 ns, along with experimental data color-coded by the approximate $\tau_{pw}$ used in each measurement. The model shows good agreement with the experimental data. The experimental reset energies are generally higher than the trend lines, primarily due to unintentional energy losses through contacts, which are not accounted for in the simplified model. **Figure 15b** illustrates the impact of TBR on reset energy, showing results for TBR = 0, 1, 10 and 100 m$^2$KGW$^{-1}$, where the TBR = 0 case corresponds to the analytical calculation and the remaining cases to finite element simulations.

**Figure 15c** estimates power consumption vs. contact diameter for different TBRs under a 10 ns reset pulse, with experimental data overlaid. The power is estimated as $P = E_{\text{min}}/\tau_{\text{pw}}$, where $E_{\text{min}}$ is the minimum reset energy per volume discussed above. Again, the model aligns well with experimental results. Finally, in **Figure 15d** we estimate the lower-limit reset current from $I = P/V$, where $V$ is a constant applied voltage (0.5, 1 or 2 V) for simplicity. We assume a GST thickness $t$ = 10 nm. Experimental reset current values from [7], [12], [33], [41], [48], [114] are plotted for comparison. Notably, the record-low reset current reported to date [40] remains at least three orders of magnitude higher than the adiabatic reset limit.

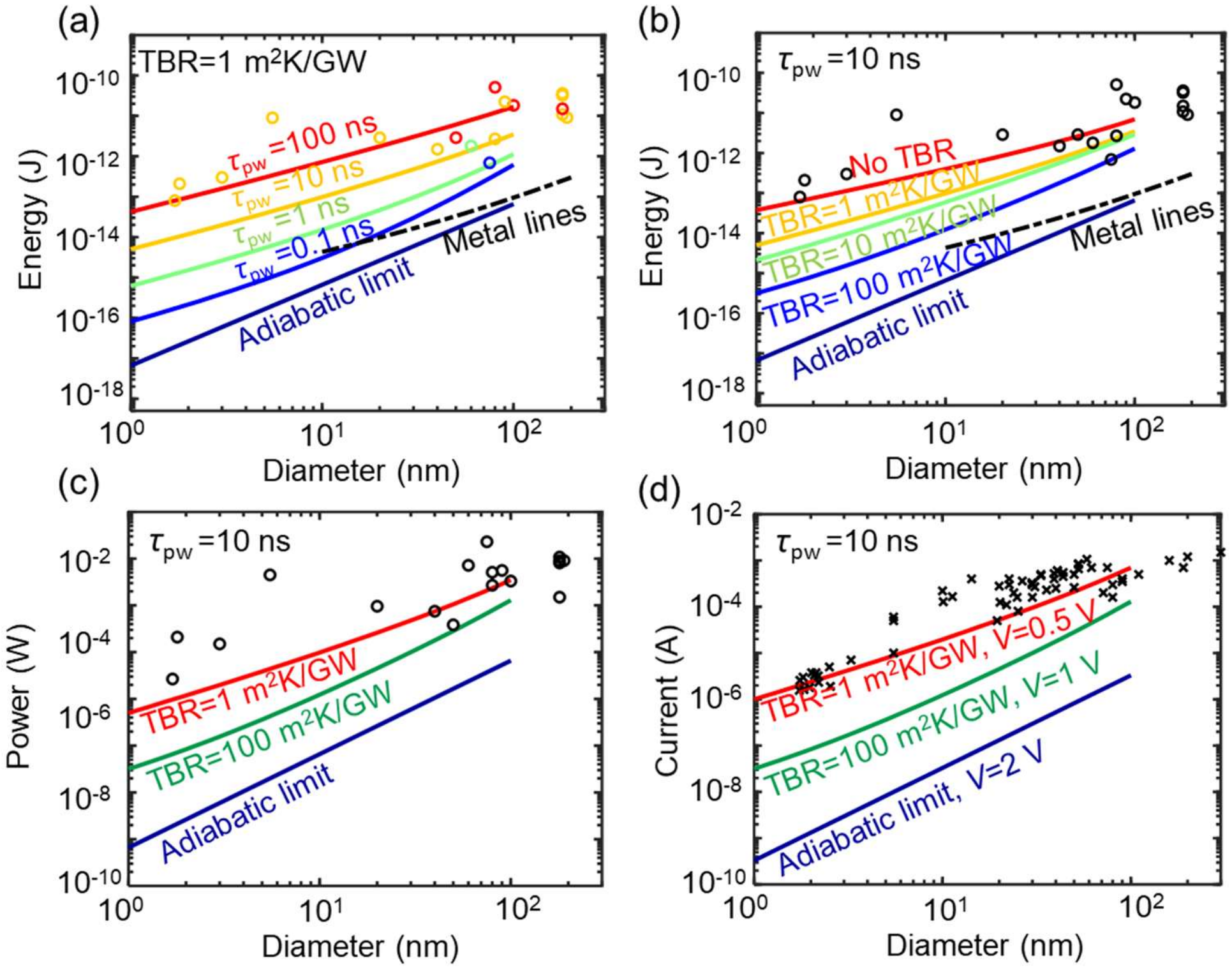

*Figure* 15 **PCM reset energy, power and current scaling.** (a) Reset energy vs. diameter for varying pulse width ($\tau_{\text{pw}}$) with fixed TBR = 1 m$^2$K/GW. The adiabatic limit used is the disk model for GST described in the text (dark blue line); metal lines' energy to charge interconnects of a single bit in 1k×1k array (dashed black line); rest of lines are based on simulations of the simplified sphere model shown in **Figure 8b**; circle markers represent experimental data points from literature and are colored based on their approximate $\tau_{\text{pw}}$. We assume the sphere model represents sufficiently practical devices with geometries up to $D \approx 100$ nm. (b) Reset energy vs. diameter for varying TBR for $\tau_{\text{pw}}$ = 10 ns. The red line represents an analytical calculation of the energy assuming zero TBR, as detailed in Supporting Information Section S6; all other curves are obtained from finite element simulations. The rest of the details are the same as (a). (c) Power vs. diameter with fixed $\tau_{pw}$ = 10 ns, assuming $P = E/\tau_{pw}$, with experimental data overlaid (circle markers). (d) Reset current vs. diameter with fixed $\tau_{pw}$ = 10 ns, assuming $I = P/V$. Cross markers represent experimental current measurements reported in the literature *[7], [12], [33], [41], [48], [114]*.

## Conclusions and Outlook

We examined recent improvements in reducing the energy consumption of phase-change memory, as well as its fundamental limits. These improvements have been primarily driven by two strategies: reducing critical device dimensions and enhancing heat confinement. Smaller dimensions reduce energy consumption by minimizing both the contact area and the volume of phase-change material that must undergo switching. Notable progress was achieved using carbon nanotube electrodes, which reduce the contact area to just a few nanometers. Graphene edge-contact electrodes and filamentary devices have also shown promise for further energy reduction through scaling. This advantage is relatively unique to PCM, as other emerging non-volatile memory (NVM) technologies like resistive RAM, being filamentary in nature, do not benefit similarly from reductions in contact diameter. However, for sub-100 nm PCM devices, the electrical and thermal behavior of contacts and interfaces becomes increasingly critical. This underscores the need to understand and engineer these interfaces for improved switching energy efficiency. Recent progress with PCM superlattice (SL) devices, featuring multiple van der Waals-like internal interfaces, highlights the potential of interface engineering to significantly lower switching energy.

We also introduced the adiabatic energy limit of PCM, *i.e.* the theoretical minimum energy required for a phase transition without any heat loss to the surrounding environment. Although current devices operate well above this idealized limit, practical progress has been made by increasing thermal boundary resistance (*e.g.,* with interfacial layers such as $Ta_2O_5$ or $MoS_2$) and by shortening programming pulse widths to the sub-nanosecond regime. These strategies limit heat spreading, improving energy efficiency and bringing reset operation closer to the adiabatic limit. However, set operations still require longer pulses, which keeps their total energy consumption on par with reset or even higher, despite requiring lower switching power. Efforts to reduce set times continue, but they remain roughly an order of magnitude longer than reset times today.

In conclusion, recent breakthroughs have significantly improved the energy efficiency of PCM, bringing it closer to the interconnect energy limit, a practical target for energy performance in memory devices and arrays. From a scientific and technological perspective, further work is needed to understand the ultimate scaling limits of PCM thickness and to investigate the thermal and electrical behavior at elevated temperatures, particularly for contacts and interfaces.

In terms of broader applications, the Ovonic threshold switch (OTS) [14], also based on phase-change materials but exhibiting volatile switching behavior, is a leading candidate for memory selectors [115], [116]. While OTS devices operate below the melting temperature, they are still thermally-activated around the crystallization temperature, meaning that many of the energy reduction strategies discussed in this review are also applicable to OTS and to the emerging selector-only memory architectures [117], [118], [119], [120], [121].

A commercially available PCM product, known as 3D XPoint, was introduced by Intel and Micron in 2015 [122]. Although it featured relatively low write latency and high endurance, the product line was discontinued by 2022. PCM still awaits its "killer application" to unlock its full market potential, with neuromorphic [123] and in-memory computing [124] emerging as the most promising use cases. With the energy improvements discussed in this review, PCM performance now aligns well with the energy demands of neuromorphic hardware. Compared to other NVM technologies, PCM offers an attractive combination of low power consumption, fast switching speeds in the nanosecond regime [125], and a high on/off resistance ratio. However, several critical challenges remain, including improving multi-level cell behavior [126], reducing variability and noise [127], and mitigating resistance drift [128], [129]. Continued progress in addressing these bottlenecks will be key to unlocking the full potential of PCM and securing its role in future memory and computing technologies.

## Acknowledgements

This work was supported, in part, by the Israel Science Foundation (ISF) via Grant No. 1179/20, as well as ISF No. 3582/21 (joint NSFC-ISF research program), by the Russell Berrie Nanotechnology Institute (RBNI), Technion. Stanford authors acknowledge support from PRISM, a JUMP 2.0 center supported by the Semiconductor Research Corporation (SRC) and DARPA, and from the Stanford Non-Volatile Memory Technology Research Initiative (NMTRI). The authors thank Ilya Karpov, Divya Somvanshi, Guy Cohen, Andrea Leonardo Lacaita, and Ilan Riess for fruitful discussions.

## References

[1] M. M. Sabry Aly *et al.*, “Energy-Efficient Abundant-Data Computing: The N3XT 1,000×,” *IEEE Computer*, vol. 48, no. 24, 2015.

[2] M. Zakarya, I. U. Rahman, and A. A. Khan, “Energy Crisis, Global Warming & IT Industry: Can the IT Professionals make it better Some Day? A Review:,” in *International Conference on Emerging Technologies*, IEEE, 2012.

[3] D. J. Brown and C. Reams, “Toward energy-efficient computing,” *Communications of the ACM*, vol. 53, no. 3, pp. 50–58, Mar. 2010, doi: 10.1145/1666420.1666438.

[4] E. Masanet, A. Shehabi, N. Lei, S. Smith, and J. Koomey, “Recalibrating global data center energy-use estimates,” *Science*, vol. 367, no. 6481, Mar. 2020, doi: 10.1007/s12053-019-09809-8.

[5] A. S. G. Andrae, “Prediction Studies of Electricity Use of Global Computing in 2030,” *International Journal of Science and Engineering Investigations*, vol. 8, 2019, [Online]. Available: www.IJSEI.com

[6] T. Mastelic, A. Oleksiak, H. Claussen, I. Brandic, J. M. Pierson, and A. V. Vasilakos, “Cloud computing: Survey on energy efficiency,” *ACM Computing Surveys*, vol. 47, no. 2, Jan. 2015, doi: 10.1145/2656204.

[7] H. S. P. Wong *et al.*, “Phase change memory,” in *Proceedings of the IEEE*, 2010, pp. 2201–2227. doi: 10.1109/JPROC.2010.2070050.

[8] H. S. P. Wong *et al.*, “Metal-oxide RRAM,” in *Proceedings of the IEEE*, Institute of Electrical and Electronics Engineers Inc., 2012, pp. 1951–1970. doi: 10.1109/JPROC.2012.2190369.

[9] H. Abbas, J. Li, and D. Ang, “Conductive Bridge Random Access Memory (CBRAM): Challenges and Opportunities for Memory and Neuromorphic Computing Applications,” *Micromachines*, vol. 13, no. 5, p. 725, Apr. 2022, doi: 10.3390/mi13050725.

[10] A. V. Khvalkovskiy *et al.*, “Erratum: Basic principles of STT-MRAM cell operation in memory arrays (Journal of Physics D: Applied Physics (2013) 46 (074001)),” *Journal of Physics D: Applied Physics*, vol. 46, no. 13, Apr. 2013, doi: 10.1088/0022-3727/46/13/139601.

[11] B. Hoffer, N. Wainstein, C. M. Neumann, E. Pop, E. Yalon, and S. Kvatinsky, “Stateful Logic Using Phase Change Memory,” *IEEE Journal on Exploratory Solid-State Computational Devices and Circuits*, vol. 8, no. 2, pp. 77–83, Dec. 2022, doi: 10.1109/JXCDC.2022.3219731.

[12] S. Raoux, F. Xiong, M. Wuttig, and E. Pop, “Phase change materials and phase change memory,” *MRS Bulletin*, vol. 39, no. 8, pp. 703–710, Aug. 2014, doi: 10.1557/mrs.2014.139.

[13] S. A. Kozyukhin, P. I. Lazarenko, A. I. Popov, and I. L. Eremenko, “Phase change memory materials and their applications,” *Russian Chemical Reviews*, vol. 91, no. 9, pp. 1–37, 2022, doi: 10.1070/rcr5033.

[14] S. R. Ovshinsky, “Reversible Electrical Switching Phenomena in Disordered Structures,” *Physical Review Letters*, vol. 21, 1968.

[15] N. Yamada, “Origin, secret, and application of the ideal phase-change material GeSbTe,” *Physica Status Solidi (b)*, vol. 249, no. 10, pp. 1837–1842, Oct. 2012, doi: 10.1002/pssb.201200618.

[16] A. L. Lacaita and A. Redaelli, "The race of phase change memories to nanoscale storage and applications," *Microelectronic Engineering*, vol. 109, pp. 351–356, 2013, doi: 10.1016/j.mee.2013.02.105.
[17] S. W. Fong, C. M. Neumann, and H. S. P. Wong, "Phase-Change Memory - Towards a Storage-Class Memory," *IEEE Transactions on Electron Devices*, vol. 64, no. 11, pp. 4374–4385, Nov. 2017, doi: 10.1109/TED.2017.2746342.
[18] S. Raoux *et al.*, "Direct observation of amorphous to crystalline phase transitions in nanoparticle arrays of phase change materials," *Journal of Applied Physics*, vol. 102, no. 9, 2007, doi: 10.1063/1.2801000.
[19] B. Liu, T. Zhang, J. Xia, Z. Song, S. Feng, and B. Chen, "Nitrogen-implanted Ge2Sb2Te5 film used as multilevel storage media for phase change random access memory," *Semiconductor Science and Technology*, vol. 19, no. 6, Jun. 2004, doi: 10.1088/0268-1242/19/6/L01.
[20] T. Nirschl *et al.*, "Write Strategies for 2 and 4-bit Multi-Level Phase-Change Memory," in *IEEE International Electron Devices Meeting (IEDM)*, 2007.
[21] D. Kuzum, S. Yu, and H. S. Philip Wong, "Synaptic electronics: Materials, devices and applications," *Nanotechnology*, vol. 24, no. 38, Sep. 2013, doi: 10.1088/0957-4484/24/38/382001.
[22] D. Kuzum, R. G. D. Jeyasingh, B. Lee, and H. S. P. Wong, "Nanoelectronic programmable synapses based on phase change materials for brain-inspired computing," *Nano Letters*, vol. 12, no. 5, pp. 2179–2186, May 2012, doi: 10.1021/nl201040y.
[23] B. L. Jackson *et al.*, "Nanoscale electronic synapses using phase change devices," *ACM Journal on Emerging Technologies in Computing Systems*, vol. 9, no. 2, 2013, doi: 10.1145/2463585.2463588.
[24] G. W. Burr *et al.*, "Experimental Demonstration and Tolerancing of a Large-Scale Neural Network (165 000 Synapses) Using Phase-Change Memory as the Synaptic Weight Element," *IEEE Transactions on Electron Devices (TED)*, vol. 62, no. 11, pp. 3498–3507, Jul. 2015, doi: 10.1109/TED.2015.2439635.
[25] S. B. Eryilmaz *et al.*, "Experimental demonstration of array-level learning with phase change synaptic devices," in *2013 IEEE International Electron Devices Meeting*, Washington, DC, USA: IEEE, Dec. 2013, p. 25.5.1-25.5.4. doi: 10.1109/IEDM.2013.6724691.
[26] S. B. Eryilmaz, S. Joshi, E. Neftci, W. Wan, G. Cauwenberghs, and H.-S. P. Wong, "Neuromorphic architectures with electronic synapses," in *2016 17th International Symposium on Quality Electronic Design (ISQED)*, Santa Clara, CA, USA: IEEE, Mar. 2016, pp. 118–123. doi: 10.1109/ISQED.2016.7479186.
[27] G. W. Burr *et al.*, "Neuromorphic computing using non-volatile memory," *Advances in Physics: X*, vol. 2, no. 1, pp. 89–124, 2017, doi: 10.1080/23746149.2016.1259585.
[28] T. Tuma, A. Pantazi, M. Le Gallo, A. Sebastian, and E. Eleftheriou, "Stochastic phase-change neurons," *Nature Nanotechnology*, vol. 11, no. 8, pp. 693–699, Aug. 2016, doi: 10.1038/nnano.2016.70.
[29] C. Grezes *et al.*, "Ultra-low switching energy and scaling in electric-field-controlled nanoscale magnetic tunnel junctions with high resistance-area product," *Applied Physics Letters*, vol. 108, no. 1, p. 012403, Jan. 2016, doi: 10.1063/1.4939446.

[30] W. Wang *et al.*, “Enabling universal memory by overcoming the contradictory speed and stability nature of phase-change materials,” *Scientific Reports*, vol. 2, 2012, doi: 10.1038/srep00360.
[31] S. B. Kim and H. S. P. Wong, “Analysis of temperature in phase change memory scaling,” *IEEE Electron Device Letters*, vol. 28, no. 8, pp. 697–699, Aug. 2007, doi: 10.1109/LED.2007.901347.
[32] J. Liu, B. Yu, and M. P. Anantram, “Scaling analysis of nanowire phase-change memory,” *IEEE Electron Device Letters*, vol. 32, no. 10, pp. 1340–1342, Oct. 2011, doi: 10.1109/LED.2011.2162390.
[33] A. Pirovano, A. L. Lacaita, A. Benvenuti, F. Pellizzer, S. Hudgens, and R. Bez, “Scaling Analysis of Phase-Change Memory Technology,” in *Tech. Dig. -Int. Electron Devices Meetings*, 2003.
[34] D. Krebs, S. Raoux, C. T. Rettner, G. W. Burr, M. Salinga, and M. Wuttig, “Threshold field of phase change memory materials measured using phase change bridge devices,” *Applied Physics Letters*, vol. 95, no. 8, 2009, doi: 10.1063/1.3210792.
[35] D. D. Yu, S. Brittman, J. S. Lee, A. L. Falk, and P. Hongkun, “Minimum voltage for threshold switching in nanoscale Phase-change memory,” *Nano Letters*, vol. 8, no. 10, pp. 3429–3433, Oct. 2008, doi: 10.1021/nl802261s.
[36] Q. Zheng, Y. Wang, and J. Zhu, “Nanoscale phase-change materials and devices,” *Journal of Physics D: Applied Physics*, vol. 50, no. 24, May 2017, doi: 10.1088/1361-6463/aa70b0.
[37] G. W. Burr *et al.*, “Phase change memory technology,” *Journal of Vacuum Science & Technology B, Nanotechnology and Microelectronics: Materials, Processing, Measurement, and Phenomena*, vol. 28, no. 2, pp. 223–262, Mar. 2010, doi: 10.1116/1.3301579.
[38] H.-S. P. Wong , S. Qin, Y. C. Shen, M. Shi, J. Sohn, M. Tung, W. Wan, X. Wu, Y. Wu, S. Yu, Z. Yu, X. Zheng *et al.*, “Stanford Memory Trends,” https://nano.stanford.edu/stanford-memory-trends.
[39] E. C. Ahn, H.-S. P. Wong, and E. Pop, “Carbon nanomaterials for non-volatile memories,” *Nat Rev Mater*, vol. 3, no. 3, p. 18009, Mar. 2018, doi: 10.1038/natrevmats.2018.9.
[40] F. Xiong, A. D. Liao, D. Estrada, and E. Pop, “Low-Power Switching of Phase-Change Materials with Carbon Nanotube Electrodes,” *Science*, vol. 332, no. 6029, pp. 568–570, Apr. 2011, doi: 10.1126/science.1201938.
[41] F. Xiong *et al.*, “Self-aligned nanotube-nanowire phase change memory,” *Nano Letters*, vol. 13, no. 2, pp. 464–469, Feb. 2013, doi: 10.1021/nl3038097.
[42] R. Waser, *Nanoelectronics and Information Technology: Advanced electronic Materials and Novel Devices*. Wiley, 2012.
[43] J. Liang, R. G. D. Jeyasingh, H. Y. Chen, and H. S. P. Wong, “An ultra-low reset current cross-point phase change memory with carbon nanotube electrodes,” *IEEE Transactions on Electron Devices*, vol. 59, no. 4, pp. 1155–1163, Apr. 2012, doi: 10.1109/TED.2012.2184542.
[44] A. Daus *et al.*, “High-performance flexible nanoscale transistors based on transition metal dichalcogenides,” *Nat Electron*, vol. 4, no. 7, pp. 495–501, Jun. 2021, doi: 10.1038/s41928-021-00598-6.
[45] A. Behnam *et al.*, “Nanoscale phase change memory with graphene ribbon electrodes,” *Applied Physics Letters*, vol. 107, no. 12, Sep. 2015, doi: 10.1063/1.4931491.

[46] F. Xiong *et al.*, “Towards Ultimate Scaling Limits of Phase-Change Memory,” in *IEEE International Electron Devices Meeting (IEDM)*, 2016, p. 4.1.1-4.1.4. Accessed: Mar. 17, 2024. [Online]. Available: 10.1109/IEDM.2016.7838342
[47] C. M. Neumann, K. L. Okabe, E. Yalon, R. W. Grady, H. S. P. Wong, and E. Pop, “Engineering thermal and electrical interface properties of phase change memory with monolayer MoS 2,” *Applied Physics Letters*, vol. 114, no. 8, Feb. 2019, doi: 10.1063/1.5080959.
[48] S.-O. Park *et al.*, “Phase-change memory via a phase-changeable self-confined nano-filament,” *Nature*, Apr. 2024, doi: 10.1038/s41586-024-07230-5.
[49] M. H. Lee and Y. C. Chen, “Structure for phase change memory and the method of forming same,” 2005
[50] R. E. Scheuerlein and S. B. Herner, “Non-Volatile Memory Cell Comprising a Dielectric Layer and a Phase Change Material in Series,” Jul. 21, 2005
[51] E. Yalon *et al.*, “Energy Dissipation in Monolayer $MoS_2$ Electronics,” *Nano Lett.*, vol. 17, no. 6, pp. 3429–3433, Jun. 2017, doi: 10.1021/acs.nanolett.7b00252.
[52] E. Yalon *et al.*, “Temperature-Dependent Thermal Boundary Conductance of Monolayer $MoS_2$ by Raman Thermometry,” *ACS Appl. Mater. Interfaces*, vol. 9, no. 49, pp. 43013–43020, Dec. 2017, doi: 10.1021/acsami.7b11641.
[53] E. Pop, V. Varshney, and A. K. Roy, “Thermal properties of graphene: Fundamentals and applications,” *MRS Bull.*, vol. 37, no. 12, pp. 1273–1281, Dec. 2012, doi: 10.1557/mrs.2012.203.
[54] S. Raoux *et al.*, “Scaling properties of phase change materials,” in *2007 Non-Volatile Memory Technology Symposium*, Albuquerque, NM: IEEE, Nov. 2007, pp. 30–35. doi: 10.1109/NVMT.2007.4389940.
[55] S. Raoux, J. L. Jordan-Sweet, and A. J. Kellock, “Crystallization properties of ultrathin phase change films,” *Journal of Applied Physics*, vol. 103, no. 11, 2008, doi: 10.1063/1.2938076.
[56] M. A. Caldwell, S. Raoux, R. Y. Wang, H. S. Philip Wong, and D. J. Milliron, “Synthesis and size-dependent crystallization of colloidal germanium telluride nanoparticles,” *Journal of Materials Chemistry*, vol. 20, no. 7, pp. 1285–1291, 2010, doi: 10.1039/b917024c.
[57] H. Hayat, K. Kohary, and C. D. Wright, “Can conventional phase-change memory devices be scaled down to single-nanometre dimensions?,” *Nanotechnology*, vol. 28, no. 3, p. 035202, Jan. 2017, doi: 10.1088/1361-6528/28/3/035202.
[58] B. Kersting and M. Salinga, “Exploiting nanoscale effects in phase change memories,” *Faraday Discuss.*, vol. 213, pp. 357–370, 2019, doi: 10.1039/C8FD00119G.
[59] J. Liu and M. P. Anantram, “Low-bias electron transport properties of germanium telluride ultrathin films,” *Journal of Applied Physics*, vol. 113, no. 6, Feb. 2013, doi: 10.1063/1.4790801.
[60] R.-G. Nir-Harwood *et al.*, “Drift of Schottky Barrier Height in Phase Change Materials,” *ACS Nano*, Mar. 2024, doi: 10.1021/acsnano.3c11019.
[61] S. Shindo, Y. Sutou, J. Koike, Y. Saito, and Y. H. Song, “Contact resistivity of amorphous and crystalline GeCu2Te3 to W electrode for phase change random access memory,” *Materials Science in Semiconductor Processing*, vol. 47, pp. 1–6, Jun. 2016, doi: 10.1016/j.mssp.2016.02.006.

[62] R. Huang *et al.*, "Contact resistance measurement of Ge 2Sb 2Te 5 phase change material to TiN electrode by spacer etched nanowire," *Semiconductor Science and Technology*, vol. 29, no. 9, Sep. 2014, doi: 10.1088/0268-1242/29/9/095003.
[63] D. Roy, M. A. A. Zandt, and R. A. M. Wolters, "Specific contact resistance of phase change materials to metal electrodes," *IEEE Electron Device Letters*, vol. 31, no. 11, pp. 1293–1295, Nov. 2010, doi: 10.1109/LED.2010.2066256.
[64] G. L. Pollack, "Kapitza Resistance," *Rev. Mod. Phys.*, vol. 41, no. 1, pp. 48–81, Jan. 1969, doi: 10.1103/RevModPhys.41.48.
[65] E. T. Swartz and R. O. Pohl, "Thermal boundary resistance," *Rev. Mod. Phys.*, vol. 61, no. 3, pp. 605–668, Jul. 1989, doi: 10.1103/RevModPhys.61.605.
[66] M. Boniardi *et al.*, "Optimization Metrics for Phase Change Memory (PCM) Cell Architectures," in *IEEE Electron Devices Meeting (IEDM)*, 2014.
[67] K. Ding *et al.*, "Phase-change heterostructure enables ultralow noise and drift for memory operation," *Science*, vol. 366, no. 6462, pp. 210–215, Oct. 2019, doi: 10.1126/science.aay0291.
[68] E. K. Chua, R. Zhao, L. P. Shi, T. C. Chong, T. E. Schlesinger, and J. A. Bain, "Effect of metals and annealing on specific contact resistivity of GeTe/metal contacts," *Applied Physics Letters*, vol. 101, no. 1, Jul. 2012, doi: 10.1063/1.4732787.
[69] S. Deshmukh *et al.*, "Temperature-Dependent Contact Resistance to Nonvolatile Memory Materials," *IEEE Transactions on Electron Devices*, vol. 66, no. 9, pp. 3816–3821, Sep. 2019, doi: 10.1109/TED.2019.2929736.
[70] K. Cil *et al.*, "Electrical resistivity of liquid Ge2Sb2Te5 based on thin-film and nanoscale device measurements," *IEEE Transactions on Electron Devices*, vol. 60, no. 1, pp. 433–437, 2013, doi: 10.1109/TED.2012.2228273.
[71] S. M. Sze, Y. Li, and K. K. Ng, *Physics of Semiconductor Devices*. Wiley-Interscience, 2006.
[72] J. Lee, E. Bozorg-Grayeli, S. Kim, M. Asheghi, H. S. Philip Wong, and K. E. Goodson, "Phonon and electron transport through Ge2Sb2Te 5 films and interfaces bounded by metals," *Applied Physics Letters*, vol. 102, no. 19, May 2013, doi: 10.1063/1.4807141.
[73] J. P. Reifenberg, D. L. Kencke, and K. E. Goodson, "The impact of thermal boundary resistance in phase-change memory devices," *IEEE Electron Device Letters*, vol. 29, no. 10, pp. 1112–1114, 2008, doi: 10.1109/LED.2008.2003012.
[74] I. R. Chen and E. Pop, "Compact thermal model for vertical nanowire phase-change memory cells," *IEEE Transactions on Electron Devices*, vol. 56, no. 7, pp. 1523–1528, 2009, doi: 10.1109/TED.2009.2021364.
[75] A. I. Khan *et al.*, "Two-Fold Reduction of Switching Current Density in Phase Change Memory Using $Bi_2Te_3$ Thermoelectric Interfacial Layer," *IEEE Electron Device Lett.*, vol. 41, no. 11, pp. 1657–1660, Nov. 2020, doi: 10.1109/LED.2020.3028271.
[76] J. Lee, M. Asheghi, and K. E. Goodson, "Impact of thermoelectric phenomena on phase-change memory performance metrics and scaling," *Nanotechnology*, vol. 23, no. 20, May 2012, doi: 10.1088/0957-4484/23/20/205201.
[77] I. Friedrich, V. Weidenhof, W. Njoroge, P. Franz, and M. Wuttig, "Structural transformations of [formula omitted] films studied by electrical resistance measurements," *Journal of Applied Physics*, vol. 87, no. 9, pp. 4130–4134, May 2000, doi: 10.1063/1.373041.

[78] D. T. Castro *et al.*, "Evidence of the Thermo-Electric Thomson Effect and Influence on the Program Conditions and Cell Optimization in Phase-Change Memory Cells," 2007.
[79] D. S. Suh *et al.*, "Thermoelectric heating of Ge2Sb2 Te5 in phase change memory devices," *Applied Physics Letters*, vol. 96, no. 12, 2010, doi: 10.1063/1.3259649.
[80] J. Lee, T. Kodama, Y. Won, M. Asheghi, and K. E. Goodson, "Phase purity and the thermoelectric properties of Ge 2Sb 2Te 5 films down to 25 nm thickness," *Journal of Applied Physics*, vol. 112, no. 1, Jul. 2012, doi: 10.1063/1.4731252.
[81] K. L. Grosse, E. Pop, and W. P. King, "Nanometer-scale temperature imaging for independent observation of Joule and Peltier effects in phase change memory devices," *Review of Scientific Instruments*, vol. 85, no. 9, Sep. 2014, doi: 10.1063/1.4895715.
[82] K. L. Grosse, F. Xiong, S. Hong, W. P. King, and E. Pop, "Direct observation of nanometer-scale Joule and Peltier effects in phase change memory devices," *Applied Physics Letters*, vol. 102, no. 19, May 2013, doi: 10.1063/1.4803172.
[83] E. Yalon *et al.*, "Spatially Resolved Thermometry of Resistive Memory Devices," *Scientific Reports*, vol. 7, no. 1, Dec. 2017, doi: 10.1038/s41598-017-14498-3.
[84] J. P. Reifenberg *et al.*, "Thermal boundary resistance measurements for phase-change memory devices," *IEEE Electron Device Letters*, vol. 31, no. 1, pp. 56–58, Jan. 2010, doi: 10.1109/LED.2009.2035139.
[85] J. Chen, X. Xu, J. Zhou, and B. Li, "Interfacial thermal resistance: Past, present, and future," *Rev. Mod. Phys.*, vol. 94, no. 2, p. 025002, Apr. 2022, doi: 10.1103/RevModPhys.94.025002.
[86] X. Wu *et al.*, "Novel nanocomposite-superlattices for low energy and high stability nanoscale phase-change memory," *Nature Communications*, vol. 15, no. 1, Dec. 2024, doi: 10.1038/s41467-023-42792-4.
[87] A. I. Khan *et al.*, "Electro-Thermal Confinement Enables Improved Superlattice Phase Change Memory," *IEEE Electron Device Letters*, vol. 43, no. 2, pp. 204–207, Feb. 2022, doi: 10.1109/LED.2021.3133906.
[88] R. E. Simpson *et al.*, "Interfacial phase-change memory," *Nature Nanotechnology*, vol. 6, no. 8, pp. 501–505, 2011, doi: 10.1038/nnano.2011.96.
[89] H. Kwon, A. I. Khan, C. Perez, M. Asheghi, E. Pop, and K. E. Goodson, "Uncovering Thermal and Electrical Properties of $Sb_2Te_3$/GeTe Superlattice Films," *Nano Lett.*, vol. 21, no. 14, pp. 5984–5990, Jul. 2021, doi: 10.1021/acs.nanolett.1c00947.
[90] K. L. Okabe *et al.*, "Understanding the switching mechanism of interfacial phase change memory," *Journal of Applied Physics*, vol. 125, no. 18, May 2019, doi: 10.1063/1.5093907.
[91] D. Térébénec *et al.*, "Improvement of Phase-Change Memory Performance by Means of GeTe/$Sb_2Te_3$ Superlattices," *Physica Rapid Research Ltrs*, vol. 15, no. 3, p. 2000538, Mar. 2021, doi: 10.1002/pssr.202000538.
[92] J. Zhao *et al.*, "Probing the Melting Transitions in Phase-Change Superlattices via Thin Film Nanocalorimetry," *Nano Lett.*, vol. 23, no. 10, pp. 4587–4594, May 2023, doi: 10.1021/acs.nanolett.3c01049.
[93] A. I. Khan *et al.*, "Unveiling the Effect of Superlattice Interfaces and Intermixing on Phase Change Memory Performance," *Nano Letters*, vol. 22, no. 15, pp. 6285–6291, Aug. 2022, doi: 10.1021/acs.nanolett.2c01869.

[94] X. Wu *et al.*, “Understanding Interface-Controlled Resistance Drift in Superlattice Phase Change Memory,” *IEEE Electron Device Letters*, vol. 43, no. 10, pp. 1669–1672, Oct. 2022, doi: 10.1109/LED.2022.3203971.
[95] A. Intisar Khan *et al.*, “First Demonstration of Ge2Sb2Te5-Based Superlattice Phase Change Memory with Low Reset Current Density (-3 MA/cm2) and Low Resistance Drift (-0.002 at 105°C),” in *Digest of Technical Papers - Symposium on VLSI Technology*, Institute of Electrical and Electronics Engineers Inc., 2022, pp. 310–311. doi: 10.1109/VLSITechnologyandCir46769.2022.9830348.
[96] A. I. Khan *et al.*, “Energy Efficient Neuro-Inspired Phase–Change Memory Based on Ge4Sb6Te7 as a Novel Epitaxial Nanocomposite,” *Advanced Materials*, vol. 35, no. 30, Jul. 2023, doi: 10.1002/adma.202300107.
[97] C. Xu, Z. Song, B. Liu, S. Feng, and B. Chen, “Lower current operation of phase change memory cell with a thin Ti O2 layer,” *Applied Physics Letters*, vol. 92, no. 6, 2008, doi: 10.1063/1.2841655.
[98] J. L. Battaglia *et al.*, “Thermal characterization of the SiO2-Ge2Sb 2Te5 interface from room temperature up to 400 °c,” *Journal of Applied Physics*, vol. 107, no. 4, 2010, doi: 10.1063/1.3284084.
[99] K. Stern *et al.*, “Uncovering Phase Change Memory Energy Limits by Sub-Nanosecond Probing of Power Dissipation Dynamics,” *Advanced Electronic Materials*, vol. 7, no. 8, Aug. 2021, doi: 10.1002/aelm.202100217.
[100] Y. Matsui *et al.*, “Ta2O5 Interfacial Layer between GST and W Plug enabling Low Power Operation of Phase Change Memories,” in *International Electron Devices Meeting (IEDM)*, 2006.
[101] S. W. Fong, C. M. Neumann, E. Yalon, M. M. Rojo, E. Pop, and H. S. P. Wong, “Dual-Layer Dielectric Stack for Thermally Isolated Low-Energy Phase-Change Memory,” *IEEE Transactions on Electron Devices*, vol. 64, no. 11, pp. 4496–4502, Nov. 2017, doi: 10.1109/TED.2017.2756071.
[102] A. Intisar Khan *et al.*, “PHASE-CHANGE MEMORY Ultralow-switching current density multilevel phase-change memory on a flexible substrate.” [Online]. Available: https://www.science.org
[103] G. M. Cohen *et al.*, “Low RESET Current Mushroom-Cell Phase-Change Memory Using Fiber-Textured Homostructure GeSbTe on Highly Oriented Seed Layer,” *physica status solidi (RRL) – Rapid Research Letters*, Jun. 2024, doi: 10.1002/pssr.202300426.
[104] W. J. Wang *et al.*, “Fast phase transitions induced by picosecond electrical pulses on phase change memory cells,” *Applied Physics Letters*, vol. 93, no. 4, 2008, doi: 10.1063/1.2963196.
[105] W. Wang *et al.*, “Nanoscaling of phase change memory cells for high speed memory applications,” *Japanese Journal of Applied Physics*, vol. 48, no. 4 PART 2, Apr. 2009, doi: 10.1143/JJAP.48.04C060.
[106] D. Loke *et al.*, “Breaking the Speed Limits of Phase-Change Memory,” *Science*, vol. 336, no. 6088, pp. 1561–1566, Jun. 2012, doi: 10.1126/science.1220119.
[107] K. Kohary and C. D. Wright, “Electric field induced crystallization in phase-change materials for memory applications,” *Applied Physics Letters*, vol. 98, no. 22, May 2011, doi: 10.1063/1.3595408.

[108] Y. Wang *et al.*, “Scandium doped Ge2Sb2Te5 for high-speed and low-power-consumption phase change memory,” *Applied Physics Letters*, vol. 112, no. 13, Mar. 2018, doi: 10.1063/1.5012872.
[109] M. Boniardi *et al.*, “Impact of Ge-Sb-Te compound engineering on the set operation performance in phase-change memories,” in *Solid-State Electronics*, Apr. 2011, pp. 11–16. doi: 10.1016/j.sse.2010.11.033.
[110] G. Bruns *et al.*, “Nanosecond switching in GeTe phase change memory cells,” *Applied Physics Letters*, vol. 95, no. 4, 2009, doi: 10.1063/1.3191670.
[111] F. Rao *et al.*, “PHASE-CHANGE MEMORY Reducing the stochasticity of crystal nucleation to enable subnanosecond memory writing,” 2017. [Online]. Available: https://www.science.org
[112] R. Zhao *et al.*, “650 ps SET speed in $Ge_2Sb_2Te_5$ phase change memory induced by $TiO_2$ dielectric crystal plane,” *InfoMat*, p. e12598, Jul. 2024, doi: 10.1002/inf2.12598.
[113] E. Ordan, R. G. Nir-Harwood, M. M. Dahan, Y. Keller, and E. Yalon, “Crystallization dynamics probed by transient resistance in phase change memory cells,” *Journal of Applied Physics*, vol. 135, no. 20, May 2024, doi: 10.1063/5.0202152.
[114] Y. Sasago *et al.*, “Cross-point phase change memory with 4F^2 cell size driven by low-contact-resistivity poly-Si diode,” in *Symposium on VLSI Technology*, 2009.
[115] W. C. Chien *et al.*, “A study on OTS-PCM pillar cell for 3-d stackable memory,” *IEEE Transactions on Electron Devices*, vol. 65, no. 11, pp. 5172–5179, Nov. 2018, doi: 10.1109/TED.2018.2871197.
[116] G. W. Burr *et al.*, “Access devices for 3D crosspoint memory,” *Journal of Vacuum Science & Technology B, Nanotechnology and Microelectronics: Materials, Processing, Measurement, and Phenomena*, vol. 32, no. 4, p. 040802, Jul. 2014, doi: 10.1116/1.4889999.
[117] T. Ravsher *et al.*, “Polarity-Induced Threshold Voltage Shift in Ovonic Threshold Switching Chalcogenides and the Impact of Material Composition,” *Physica Status Solidi - Rapid Research Letters*, vol. 17, no. 8, Aug. 2023, doi: 10.1002/pssr.202200417.
[118] T. Ravsher *et al.*, “Self-Rectifying Memory Cell Based on SiGeAsSe Ovonic Threshold Switch,” *IEEE Trans. Electron Devices*, vol. 70, no. 5, pp. 2276–2281, May 2023, doi: 10.1109/TED.2023.3252491.
[119] T. Ravsher *et al.*, “Comprehensive Performance and Reliability Assessment of Se-based Selector-Only Memory,” in *2024 IEEE International Reliability Physics Symposium (IRPS)*, Grapevine, TX, USA: IEEE, Apr. 2024, p. 7A.5-1-7A.5-9. doi: 10.1109/IRPS48228.2024.10529450.
[120] T. Ravsher *et al.*, “Polarity-dependent threshold voltage shift in ovonic threshold switches: Challenges and opportunities,” in *2021 IEEE International Electron Devices Meeting (IEDM)*, San Francisco, CA, USA: IEEE, Dec. 2021, p. 28.4.1-28.4.4. doi: 10.1109/IEDM19574.2021.9720649.
[121] S. Clima *et al.*, “Selector Only Memory: Exploring Atomic Mechanisms from First-Principles,” in *2024 IEEE International Electron Devices Meeting (IEDM)*, San Francisco, CA, USA: IEEE, Dec. 2024, pp. 1–4. doi: 10.1109/IEDM50854.2024.10873555.
[122] H.-Y. Cheng, F. Carta, W.-C. Chien, H.-L. Lung, and M. J. BrightSky, “3D cross-point phase-change memory for storage-class memory,” *J. Phys. D: Appl. Phys.*, vol. 52, no. 47, p. 473002, Nov. 2019, doi: 10.1088/1361-6463/ab39a0.

[123] T. Kim and S. Lee, "Evolution of Phase-Change Memory for the Storage-Class Memory and Beyond," *IEEE Trans. Electron Devices*, vol. 67, no. 4, pp. 1394–1406, Apr. 2020, doi: 10.1109/TED.2020.2964640.

[124] K. Fan *et al.*, "SpecPCM: A Low-Power PCM-Based In-Memory Computing Accelerator for Full-Stack Mass Spectrometry Analysis," *IEEE J. Explor. Solid-State Comput. Devices Circuits*, vol. 10, pp. 161–169, 2024, doi: 10.1109/JXCDC.2024.3498837.

[125] P. Jangra and M. Duhan, "Performance-based comparative study of existing and emerging non-volatile memories: a review," *J Opt*, vol. 52, no. 4, pp. 2395–2409, Dec. 2023, doi: 10.1007/s12596-022-01058-w.

[126] N. Papandreou *et al.*, "Multilevel phase-change memory," in *2010 17th IEEE International Conference on Electronics, Circuits and Systems*, Athens, Greece: IEEE, Dec. 2010, pp. 1017–1020. doi: 10.1109/ICECS.2010.5724687.

[127] G. F. Close *et al.*, "Device, circuit and system-level analysis of noise in multi-bit phase-change memory," in *IEEE International Electron Devices Meeting (IEDM)*, 2010.

[128] A. Pirovano, A. L. Lacaita, F. Pellizzer, S. A. Kostylev, A. Benvenuti, and R. Bez, "Low-field amorphous state resistance and threshold voltage drift in chalcogenide materials," *IEEE Transactions on Electron Devices (TED)*, vol. 51, no. 5, pp. 714–719, May 2004, doi: 10.1109/TED.2004.825805.

[129] J. Y. Raty *et al.*, "Aging mechanisms in amorphous phase-change materials," *Nature Communications*, vol. 6, Jun. 2015, doi: 10.1038/ncomms8467.

# Supporting Information

## Energy and Scaling Limits of Phase-Change Memory

Rivka-Galya Nir-Harwood[1], Asir I. Khan[2,3], Emanuel Ber[1], Efrat Ordan[1], Keren Stern[1], Kye L. Okabe[2], Nicolás Wainstein[1], Eilam Yalon[1*], and Eric Pop[2,4,5,6**]

[1]*Viterbi faculty of Electrical & Computer Engineering, Technion - Israel Institute of Technology, Haifa, 32000, Israel.*
[2]*Department of Electrical Engineering, Stanford University, Stanford, CA 94305, USA.*
[3]*Lawrence Berkeley National Laboratory, Berkeley, CA 94720, USA.*
[4]*Department of Materials Science & Engineering, Stanford University, Stanford, CA 94305, USA.*
[5]*Department of Applied Physics, Stanford University, Stanford, CA 94305, USA.*
[6]*Precourt Institute for Energy, Stanford University, Stanford, CA 94305, USA.*

[*]*E-mail: eilamy@technion.ac.il*
[**]*E-mail: epop@stanford.edu*

Contents

## S1. Additional Device Structures

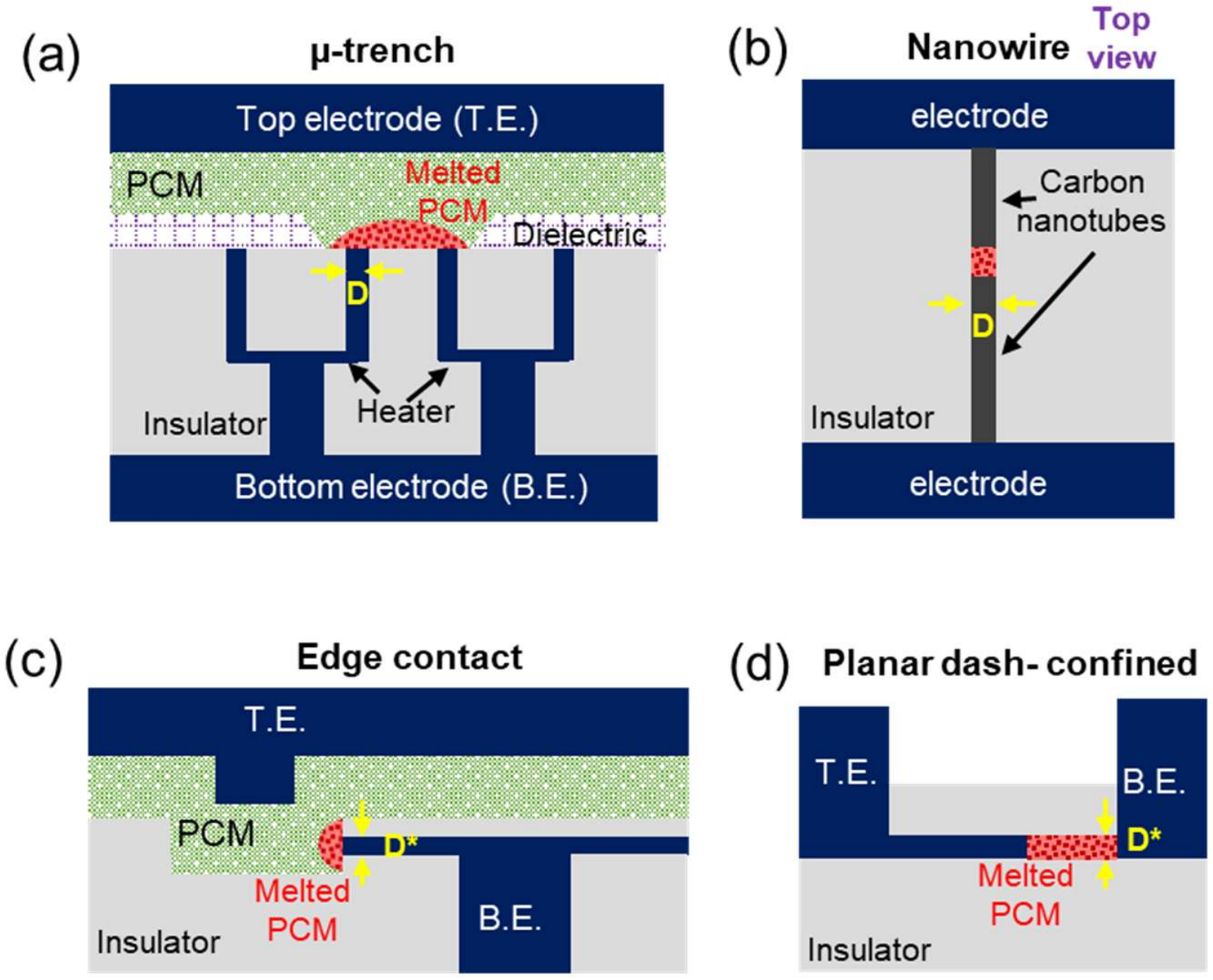

*Supporting Figure S1* **Schematic cross-sectional device structure of PCM cells.** (a) μ-trench (b) nanowire (c) edge contact, and (d) planar dash confined. The critical size of the bottom electrode or confinement is marked as "D" (diameter), and the melted PCM region during device operation is shown. For structures that do not have a symmetrical circular contact (exp. planar dash-confined and edge contact) the diameter used is the square root of the contact area and marked as "D*".

## S2. Metal Interconnect Energy Calculation

In this section we calculate the energy consumptions of the metal interconnect lines in a 1k×1k array. First, the capacitance, $C_1$, per unit length of a single metal line placed on bulk silicon, shown in Supporting Figure S2a can be estimated as [1] :

$$C_1 = \epsilon \cdot (1.15 \cdot \left(\frac{W}{H}\right) + 2.80 \cdot \left(\frac{T}{H}\right)^{0.222})$$

Where $\varepsilon$ is the dielectric constant of the insulator (in units of F/m), $W$ is the width of the metal line, $T$ is the thickness of the metal line and $H$ is the height of the metal line from the bulk silicon.

In the case of three lines, the total capacitance of the middle line per unit length, shown in Supporting Figure S2b can be empirically expressed as:

$$C_2 = C_1 + \epsilon \cdot 2 \cdot \left[0.03 \cdot \left(\frac{W}{H}\right) + 0.83 \cdot \left(\frac{T}{H}\right) - 0.07 \cdot \left(\frac{T}{H}\right)^{0.222}\right] \cdot \left(\frac{S}{H}\right)^{-1.34}$$

Where $S$ is the distance between lines. For this estimation to have an error less than 10% we must maintain $0.5 \cdot H < W, T, S < 10 \cdot H$. We assume the height of the metal line is 20nm.

The energy is calculated using:

$$E = \frac{C \cdot V^2}{2}$$

Where $V$ is the applied voltage. We assume 1 V in energy figures in main text and a 1k×1k array (*i.e.*, the metal line length is $1000 \cdot (W + S)$ in units of nm).

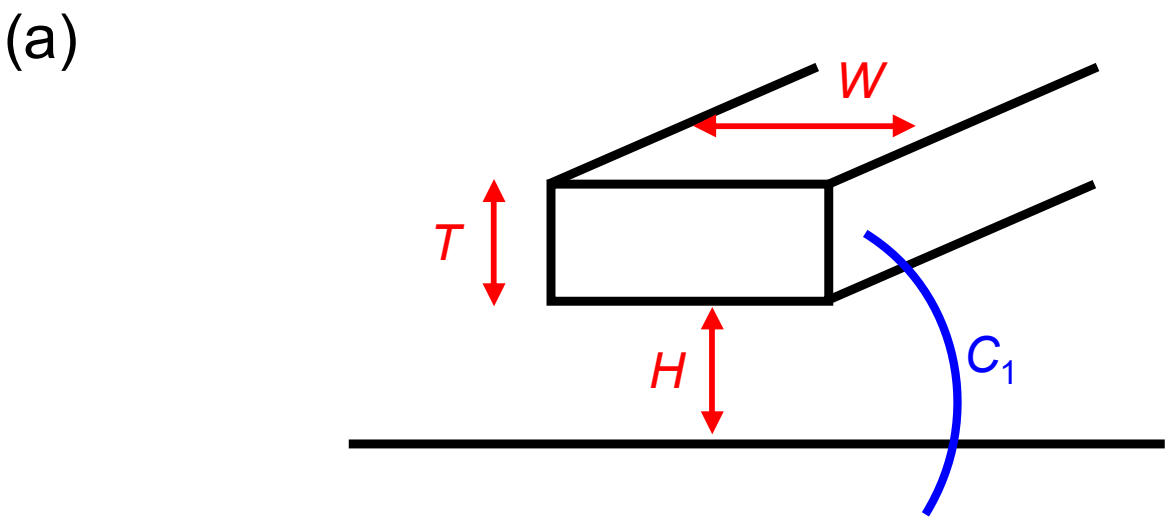

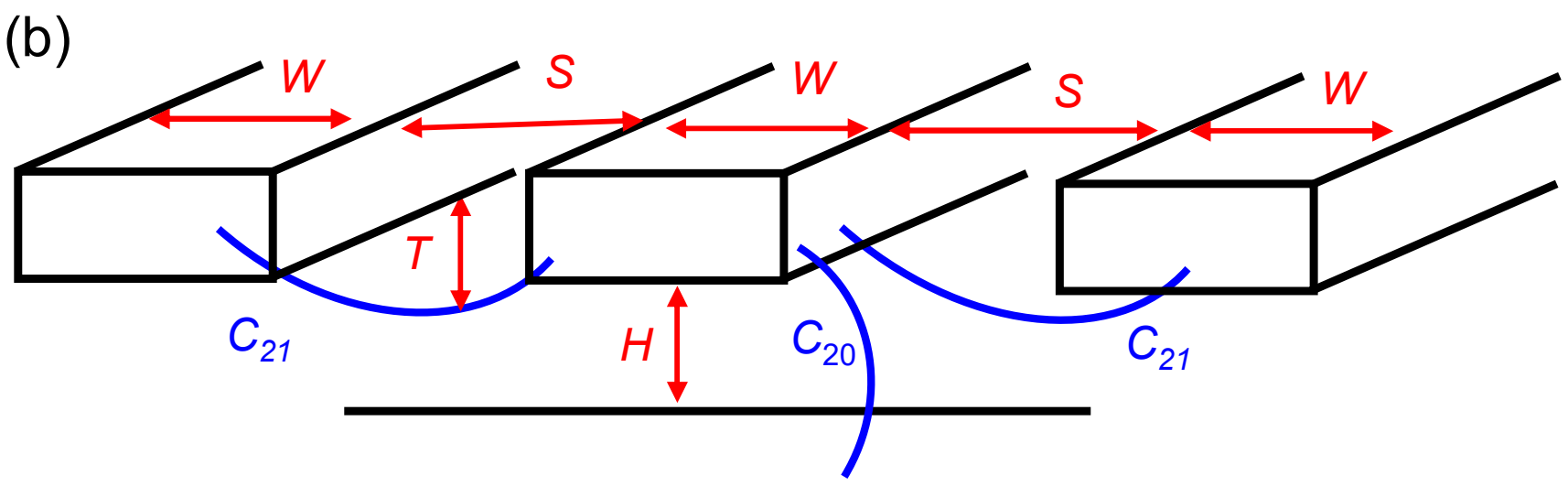

*Supporting Figure S2* **Geometry of wiring** (a) Single line on ground plane (b) three lines on ground plane

## S3. PCM Resistivity Table

*Table 1* ***PCM resistivity****. For different phase change materials and metal contacts, resistivity ρ, specific contact resistivity $\rho_c$, and characteristic length at which bulk resistance equals contact resistance γ, in both crystalline and amorphous phases. Works indicate that for 100-nm-long (lateral) or -thick (vertical) devices, in both phases, the resistance is dominated by the contacts.*

| **Contact** | **a-GST measurement strategy** | **a-GST type** | $\rho_a$ **(mΩ·cm) a-GST bulk resistivity** | $\rho_c$ **(mΩ·cm) c-GST bulk resistivity** | $\rho^C_a$ **(mΩ·cm²) a-GST contact resistivity** | $\rho^C_c$ **(mΩ·cm²) c-GST contact resistivity** | $\gamma_a$ **(µm)** $\rho^C_a/\rho_a$ | $\gamma_c$ **(µm)** $\rho^C_c/\rho_c$ |
|---|---|---|---|---|---|---|---|---|
| **Nir-Harwood** [2] **$Ge_2Sb_2Te_5$-** TiN | TLM | Ge ion implanted | $22\cdot10^3$ | 30 | 3.3 | $1.6\cdot10^{-3}$ | 1.57 | 0.53 |
| **Roy** [3] **$Ge_2Sb_2Te_5$-** TiW | Cross-bridge Kelvin resistor | As-deposited | $80.5\cdot10^4$ | 20 | 95 | $3.8\cdot10^{-3}$ (175˚ anneal) $0.2\cdot10^{-3}$ (350˚ anneal) | 1.18 | 1.90 |
| **Huang** [4] **$Ge_2Sb_2Te_5$-** TiN | Modified TLM | As-deposited | $54.5\cdot10^3$ | 34 | 64 | $80\cdot10^{-3}$ | 11.74 | 23.52 |
| **Shindo** [5] **$Ge_2Sb_2Te_5$-** W | Circular TLM | As-deposited | $22\cdot10^5$ | 48 | 32 | $14\cdot10^{-3}$ | 0.15 | 2.91 |
| **Adnane** [6] **$Ge_2Sb_2Te_5$-** W | 4-point-probe | As-deposited | $87\cdot10^4$ | 75 (fcc) 11 (hcp) | N/A | N/A | N/A | N/A |
| **Chua** [7] **GeTe-TiW** | Circular TLM | As-deposited | $25\cdot10^5$ | 0.54 | 0.6 | $14.3\cdot10^{-6}$ | <0.01 | 0.26 |
| **Savransky** [8] **$Ge_2Sb_2Te_5$-** metal | varying GST thicknesses | Melt-quenched | $10\cdot10^5$ | 20 | N/A | $0.3\cdot10^{-3}$ | N/A | 0.15 |

## S4. Thermal Boundary Resistance Table

*Table 2* ***GST thermal boundary resistance (TBR) table***

| PCM | Interface with | TBR value ($m^2K/GW$) | | | Source |
|---|---|---|---|---|---|
| | | amorphous | fcc | hcp | |
| GST | Al | 20±5 | 6±2 | 80±20 | *TDTR measurement [9] |
| GST | $Si_3N_4$ | 134±50 | 25±8 | | TDTR measurement [10] |
| GST | Pt | | 25±9 | | Raman thermometry [11] |
| GST | $SiO_2$ | | 28±8 | | Raman thermometry [11] |
| GST | $SiO_2$ | | 29±20 | | TDTR measurement [10] |
| GST | $SiO_2$ | ~50 | ~5 | ~120 | TDTR measurement [12] |
| GST | $SiO_2$ | | | ~90 | TDTR measurement at ~ 400°C [12] |
| GST | TiN | 210±80 | 24±10 | 12±10 | TDTR measurement [13] |
| GST | Al or TiN or $SiO_2$ | | | ~20 | First-principles calculations [14] |

*TDTR refers to time-domain thermo-reflectance

## S5. Minimum Energy to Melt Table

*Table 3* ***Minimum energy (per volume) to melt***

| Material | $C_s$ ($J/cm^3/K$) | $T_M$ (K) | $\Delta T_M$ (K) | $\Delta H$ ($J/cm^3$) | $Q = C_s \times \Delta T_M$ ($J/cm^3$) | $E_{min} = Q + \Delta H$ ($J/cm^3$) |
|---|---|---|---|---|---|---|
| $Ge_2Sb_2Te_5$ | 1.34 [12] | | ~640 | 90 [12], [15] | 770 | 860 |
| GeTe-$Sb_2Te_3$-Sb | 1.28 [13] | | ~600 | 420 [13] | 740 | 1160 |
| GeTe | 1.58 [11] | | ~725 | 1450 [11] | 1105 | 2555 |
| $Sb_2Te_3$ | 1.33 [16] | | ~620 | 1025 [14] | 790 | 1820 |
| $Sb_7Te_3$ | 1.39 | | ~540 | | 715 | |
| Ge | 1.71 | 1211 | ~940 | 2320 | 1465 | 3735 |
| Te | 1.25 | 723 | ~450 | 850 | 530 | 1380 |
| Sb | 1.39 | 904 | ~630 | 1080 | 850 | 1940 |

## S6. Analytical Energy Calculations for Sphere

The dimensionless temperature distribution outside a sphere of radius $R$ into an isotropic space whose properties are constant and whose initial temperature $T_i$ is constant, based on Dirichlet is [17] :

$$\phi = \frac{1}{\rho} \cdot erfc(\frac{\rho - 1}{2\sqrt{Fo}})$$

Where $\rho = r/R$, and $\mathrm{Fo} = \alpha\tau_{pw}/R^2$ ($\alpha$ is the thermal diffusivity in units of $m^2/s$) and $\tau_{pw}$ is the pulse length.

Assuming a thermal boundary resistance (TBR) of 0, the energy beyond the sphere is calculated by multiplying the dimensionless temperature distribution ($\phi$) by the melting temperature ($T_M$) and the sphere area, and integrating it between $R$ and $\infty$ as follows:

$$E_{beyond\ sphere} = C_s \cdot \int_R^{\infty} T_M \cdot 4\pi r^2 \cdot \phi(r) dr$$

This estimation is plotted in red in Figure 15b in the main text where $R$ is half the device diameter.

## S7. PCM Energy vs Diameter Table

*Table 4* ***PCM Energy vs Diameter experimental data.***

| ref | D (nm) | PCM material | Device strucutre | $E_R$ (pJ) | $V_R$ (V) | $I_R$ (µA) | $PW_R$ (ns) | $E_S$ (pJ) | $V_S$ (V) | $I_S$ (µA) | $PW_S$ (ns) |
|---|---|---|---|---|---|---|---|---|---|---|---|
| [18] | 180 | GST 225 | µ-trench | 36 | 1.5 | 600 | 40 | 20 | 1 | 200 | 100 |
| [19] | 180 | GST 225 | µ-trench | 32.4 | 2.7 | 300 | 40 | 90 | 2 | 300 | 150 |
| [20] | 180 | GST 225 | --- | 10.8 | 1.8 | 600 | 10 | 64.8 | 1.8 | 200 | 180 |
| [21] | 180 | GST doped | mushroom | 15 | 1.5 | 100 | 100 | 36 | 1.2 | 100 | 300 |
| [22] | 3 | GST 225 | nanowire | 0.3 | 3 | 5 | 20 | 0.45 | 3 | 1 | 150 |
| [23] | 1.7 | GST 225 | nanowire | 0.08 | --- | 1.6 | 30 | 0.03 | --- | 0.1 | 300 |
| [24] | 50 | SL | mushroom | 2.89 | 0.7 | 55 | 75 | 3.6 | 0.6 | 50 | 120 |
| [25] | 90 | GST 225 | µ-trench | 22.4 | 1.4 | 400 | 40 | 20 | 1 | 200 | 100 |

| [26] | 80 | GST 225 | Dash confined | 51.2 | 3.2 | 160 | 100 | 252 | 4.2 | 60 | 1000 |
|---|---|---|---|---|---|---|---|---|---|---|---|
| [27] | 100 | GST 225 | Dash confined | 18.25 | --- | 94 | 55 | --- | --- | --- | 1100 |
| [28] | 75 | GST 225 | confined | 0.69 | 12.8 | 180 | 0.3 | | | | |
| [29] | 175 | GST 225 | confined | | | | | 720 | 1.8 | 2000 | 200 |
| [30] | 40 | SL | | 1.5 | 0.375 | 200 | 20 | 0.91 | 0.65 | 40 | 35 |
| [31] | 80 | SL | mushroom | 2.7 | 0.9 | 300 | 10 | | | | |
| [31] | 190 | SL | mushroom | 9.1 | 1.3 | 700 | 10 | | | | |
| [32] | 1.8 | GST 225 | nanowire | 0.21 | 15 | 1.4 | 10 | --- | 5.28 | 0.5 | --- |
| [33] | 60 | GeTe | mushroom | 1.8 | 0.6 | 1200 | 2.5 | 7.2 | 1 | 900 | 8 |
| [34]* | 5.5 | $SiTe_x$ | mushroom | 9 | --- | 60 | 20 | 2.63 | --- | 5 | 150 |
| [35] | 100 | SL | | 6 | | | 20 | | | | |
| [36]* | 150 | GST 225 | mushroom | 20 | 0.8 | 500 | 50 | --- | --- | --- | 2000 |

*This data is not shown in the scaling trend figures in the main text (Figure 14-15) because it is filamentary and the critical dimension (D) is independent of the bottom electrode size.

[1] T. Sakurai and K. Tamaru, “Simple Formulas for Two-and Three-Dimensional Capacitances,” *Trans. Electron Devices*, vol. 30, no. 2, 1983.

[2] R.-G. Nir-Harwood *et al.*, “Drift of Schottky Barrier Height in Phase Change Materials,” *ACS Nano*, Mar. 2024, doi: 10.1021/acsnano.3c11019.

[3] D. Roy, M. A. A. Zandt, and R. A. M. Wolters, “Specific contact resistance of phase change materials to metal electrodes,” *IEEE Electron Device Lett.*, vol. 31, no. 11, pp. 1293–1295, Nov. 2010, doi: 10.1109/LED.2010.2066256.

[4] R. Huang *et al.*, “Contact resistance measurement of Ge 2Sb 2Te 5 phase change material to TiN electrode by spacer etched nanowire,” *Semicond. Sci. Technol.*, vol. 29, no. 9, Sep. 2014, doi: 10.1088/0268-1242/29/9/095003.

[5] S. Shindo, Y. Sutou, J. Koike, Y. Saito, and Y. H. Song, “Contact resistivity of amorphous and crystalline GeCu2Te3 to W electrode for phase change random access memory,” *Mater. Sci. Semicond. Process.*, vol. 47, pp. 1–6, Jun. 2016, doi: 10.1016/j.mssp.2016.02.006.

[6] L. Adnane *et al.*, “High temperature electrical resistivity and Seebeck coefficient of Ge2Sb2Te5 thin films,” *J. Appl. Phys.*, vol. 122, no. 12, Sep. 2017, doi: 10.1063/1.4996218.

[7] E. K. Chua, R. Zhao, L. P. Shi, T. C. Chong, T. E. Schlesinger, and J. A. Bain, “Effect of metals and annealing on specific contact resistivity of GeTe/metal contacts,” *Appl. Phys. Lett.*, vol. 101, no. 1, Jul. 2012, doi: 10.1063/1.4732787.

[8] S. D. Savransky and I. V. Karpov, “Investigation of SET and RESET States Resistance in Ohmic Regime for Phase-Change Memory,” 2008.

[9] J. L. Battaglia *et al.*, “Thermal resistance at Al-Ge2Sb2Te5 interface,” *Appl. Phys. Lett.*, vol. 102, no. 18, May 2013, doi: 10.1063/1.4803923.

[10] J. Lee *et al.*, “Decoupled Thermal Resistances of Phase Change Material and Their Impact on PCM Devices,” in *Thermal and Thermomechanical Phenomena in Electronic Systems (ITherm), 12th IEEE Intersociety Conference*, IEEE, 2010.

[11] E. Yalon *et al.*, “Spatially Resolved Thermometry of Resistive Memory Devices,” *Sci. Rep.*, vol. 7, no. 1, Dec. 2017, doi: 10.1038/s41598-017-14498-3.

[12] J. L. Battaglia *et al.*, “Thermal characterization of the SiO2-Ge2Sb 2Te5 interface from room temperature up to 400 °c,” *J. Appl. Phys.*, vol. 107, no. 4, 2010, doi: 10.1063/1.3284084.

[13] J. Lee, E. Bozorg-Grayeli, S. Kim, M. Asheghi, H. S. Philip Wong, and K. E. Goodson, “Phonon and electron transport through Ge2Sb2Te 5 films and interfaces bounded by metals,” *Appl. Phys. Lett.*, vol. 102, no. 19, May 2013, doi: 10.1063/1.4807141.

[14] S. Gabardi, D. Campi, and M. Bernasconi, “Ab initio calculation of thermal boundary resistance at the interface of metals with GeTe, In 3 SbTe 2 and In 2 GeTe 3 phase change compounds,” *J. Comput. Electron.*, vol. 16, no. 4, pp. 1003–1010, Dec. 2017, doi: 10.1007/s10825-017-1097-1.

[15] C. Xu, Z. Song, B. Liu, S. Feng, and B. Chen, “Lower current operation of phase change memory cell with a thin Ti O2 layer,” *Appl. Phys. Lett.*, vol. 92, no. 6, 2008, doi: 10.1063/1.2841655.

[16] D. Campi, E. Baldi, G. Graceffa, G. C. Sosso, and M. Bernasconi, “Electron-phonon interaction and thermal boundary resistance at the interfaces of Ge2Sb2Te5 with metals and dielectrics,” *J. Phys. Condens. Matter*, vol. 27, no. 17, May 2015, doi: 10.1088/0953-8984/27/17/175009.

[17] W. M. Rohsenow, J. P. (James P. ) Hartnett, and Y. I. Cho, *Handbook of heat transfer*. McGraw-Hill, 1998.

[18] F. Pellizzer *et al.*, “Novel pTrench Phase-Change Memory Cell for Embedded and Stand-Alone Non-Volatile Memory Applications.”

[19] F. Bedeschi *et al.*, “An 8Mb Demonstrator for High-Density 1.8V Phase-Change Memories,” Agrate.

[20] H. R. Oh *et al.*, “Enhanced write performance of a 64-mb phase-change random access memory,” *IEEE J. Solid-State Circuits*, vol. 41, no. 1, pp. 122–126, Jan. 2006, doi: 10.1109/JSSC.2005.859016.

[21] Y. Matsui *et al.*, "Ta2O5 Interfacial Layer between GST and W Plug enabling Low Power Operation of Phase Change Memories," in *International Electron Devices Meeting (IEDM)*, 2006.
[22] F. Xiong *et al.*, "Low-Power Switching of Phase-Change Materials with Carbon Nanotube Electrodes," *Tech. Dig. - Int. Electron Devices Meet. IEDM*, vol. 332, no. 6029, 2011, doi: 10.1109/IEDM.2006.346910.
[23] F. Xiong *et al.*, "Self-aligned nanotube-nanowire phase change memory," *Nano Lett.*, vol. 13, no. 2, pp. 464–469, Feb. 2013, doi: 10.1021/nl3038097.
[24] N. Takaura *et al.*, "55-μA GexTe1-x/Sb2Te3 superlattice topological-switching random access memory (TRAM) and study of atomic arrangement in Ge-Te and Sb-Te structures," in *IEEE International Electron Devices Meeting (IEDM)*, 2014.
[25] F. Pellizzer *et al.*, "A 90nm Phase Change Memory Technology for Stand-Alone Non-Volatile Memory Applications," 2006.
[26] Y. Sasago *et al.*, "Cross-point phase change memory with 4F^2 cell size driven by low-contact-resistivity poly-Si diode," in *Symposium on VLSI Technology*, 2009.
[27] S. W. Fong, C. M. Neumann, E. Yalon, M. M. Rojo, E. Pop, and H. S. P. Wong, "Dual-Layer Dielectric Stack for Thermally Isolated Low-Energy Phase-Change Memory," *IEEE Trans. Electron Devices*, vol. 64, no. 11, pp. 4496–4502, Nov. 2017, doi: 10.1109/TED.2017.2756071.
[28] K. Stern *et al.*, "Uncovering Phase Change Memory Energy Limits by Sub-Nanosecond Probing of Power Dissipation Dynamics," *Adv. Electron. Mater.*, vol. 7, no. 8, Aug. 2021, doi: 10.1002/aelm.202100217.
[29] E. Ordan, R. G. Nir-Harwood, M. M. Dahan, Y. Keller, and E. Yalon, "Crystallization dynamics probed by transient resistance in phase change memory cells," *J. Appl. Phys.*, vol. 135, no. 20, May 2024, doi: 10.1063/5.0202152.
[30] X. Wu *et al.*, "Novel nanocomposite-superlattices for low energy and high stability nanoscale phase-change memory," *Nat. Commun.*, vol. 15, no. 1, Dec. 2024, doi: 10.1038/s41467-023-42792-4.
[31] K. Ding *et al.*, "Phase-change heterostructure enables ultralow noise and drift for memory operation," 2019. [Online]. Available: https://www.science.org
[32] J. Liang, R. G. D. Jeyasingh, H. Y. Chen, and H. S. P. Wong, "An ultra-low reset current cross-point phase change memory with carbon nanotube electrodes," *IEEE Trans. Electron Devices*, vol. 59, no. 4, pp. 1155–1163, Apr. 2012, doi: 10.1109/TED.2012.2184542.
[33] G. Bruns *et al.*, "Nanosecond switching in GeTe phase change memory cells," *Appl. Phys. Lett.*, vol. 95, no. 4, 2009, doi: 10.1063/1.3191670.
[34] S.-O. Park *et al.*, "Phase-change memory via a phase-changeable self-confined nano-filament," *Nature*, Apr. 2024, doi: 10.1038/s41586-024-07230-5.
[35] A. I. Khan *et al.*, "Electro-Thermal Confinement Enables Improved Superlattice Phase Change Memory," *IEEE Electron Device Lett.*, vol. 43, no. 2, pp. 204–207, Feb. 2022, doi: 10.1109/LED.2021.3133906.
[36] C. M. Neumann, K. L. Okabe, E. Yalon, R. W. Grady, H. S. P. Wong, and E. Pop, "Engineering thermal and electrical interface properties of phase change memory with monolayer MoS 2," *Appl. Phys. Lett.*, vol. 114, no. 8, Feb. 2019, doi: 10.1063/1.5080959.